\documentclass[aps,prx,twocolumn,a4paper,10pt,nofootinbib,showkeys,longbibliography]{revtex4-2}
\usepackage[utf8]{inputenc}
\usepackage[T1]{fontenc}
\usepackage{lmodern}
\usepackage[english]{babel}
\usepackage{xcolor}
\usepackage[intlimits]{amsmath}
\usepackage{amssymb,amsthm}
\usepackage{mathtools}
\usepackage{graphicx}
\usepackage{textcomp}
\usepackage{enumerate}
\definecolor{green}{HTML}{00643A}
\usepackage[colorlinks=true,allcolors=green,breaklinks=true]{hyperref}
\usepackage[capitalise]{cleveref}
\usepackage{booktabs}
\usepackage[]{units}
\usepackage[]{units}
\usepackage{braket}
\usepackage{bm}
\usepackage{url}
\usepackage[normalem]{ulem}

\newcommand{\cc}{\text{c}}
\newcommand{\dd}{\text{d}}
\newcommand{\dt}{\text{d}t}
\newcommand{\kb}{k_\text{B}}
\newcommand{\kT}{k_\text{B}T}
\newcommand{\etal}{\emph{et~al.}}
\newcommand{\cf}{\emph{cf.~}}
\newcommand{\eg}{\emph{e.g.~}}
\newcommand{\ie}{\emph{i.e.}}
\newcommand{\uu}{\text{u}}
\newcommand{\Sc}{\text{S}_\text{c}}
\newcommand{\Su}{\text{S}_\text{u}}
\newcommand{\Mc}{\text{M}_\text{c}}
\newcommand{\Mu}{\text{M}_\text{u}}
\newcommand{\EM}[0]{\ensuremath{E_\mathrm{M}}}
\newcommand{\ES}[0]{\ensuremath{E_\mathrm{S}}}

\begin{document}

\title{Controlling energy delivery with bistable nanostructures}

\author{Andreas Ehrmann}
\author{Carl P. Goodrich}
\email{carl.goodrich@ist.ac.at}
\affiliation{Institute of Science and Technology Austria (ISTA), Am Campus 1, 3400 Klosterneuburg, Austria}
\date{\today}

\begin{abstract}
    Countless biological processes are fueled by energy-rich molecules like ATP and GTP that supply energy with extreme efficiency. However, designing similar energy-delivery schemes from the bottom up, essential for the development of powered nanostructures and other \emph{de novo} machinery, presents a significant challenge: how can an energy-rich structure be stable in solution yet still deliver this energy at precisely the right time? In this paper, we present a purely physical mechanism that solves this challenge, facilitating energy transfer akin to ATP hydrolysis, yet occurring between synthetic nanostructures without any biochemical interactions. This targeted energy delivery is achieved by exploiting a differentiable state-based model to balance the energy profiles that govern the structural transitions in the two nanostructures, creating a coupled relaxation pathway with minimal barriers that facilitates energy delivery. We verify the effectiveness and robustness of this mechanism through Langevin Dynamics simulations, demonstrating that a bath of the high-energy structures can systematically and repeatedly drive the target structure out of equilibrium, enabling it to perform tasks. As the mechanism operates only through explicit physical forces without any biochemistry or internal state variables, our results present generic and far-reaching design principles, setting the stage for the next generation of synthetic nanomachines. 
\end{abstract}


\maketitle

\section{Introduction}
A hallmark of life is the precise control of energy transduction -- energetically charged molecules such as ATP and GTP deliver packets of energy where and when they are needed with extreme efficiency~\cite{ross06, phil12, nels13, lipp19}. 
This targeted energy delivery powers a wide range of processes~\cite{phil12, nels13}; for example, motor proteins like Myosin and Kinesin convert the chemical energy of ATP into mechanical work to generate forces and perform tasks~\cite{cros00, fish01, cart05, ould13} such as muscle contraction and cargo transport. Moreover, such biological nanomachines can operate simultaneously and independently in part due to the precision with which energy delivery is controlled, allowing for the overwhelming complexity found within cells~\cite{brow20}. 

Despite enormous experimental~\cite{leun09, roge16, jaco24} and theoretical~\cite{juli97, zera17, good21, das23, jhav23, gart24} progress, efforts to engineer synthetic nanomachines analogous to motor proteins with this level of complexity and functionality are still in their infancy compared to their biological counterparts. Recent pioneering studies have revealed the design of mechanical catalysts~\cite{agud21, muno23}, chemically fueled molecular machines~\cite{alba22, alba23}, and fueled enzymatic behavior~\cite{chat25}. However, one of the remaining key hurdles in bio-inspired nanotechnology is the lack of a generalizable mechanism for controlled energy delivery. Although synthetic energy transduction can be achieved, for example, through external fields or active particles~\cite{brow20, aubr21, shen23}, these brute-force approaches lack the precision of ATP hydrolysis.

In this paper, we demonstrate a robust physical mechanism for controlled energy delivery between a pair of two-state nanostructures, roughly analogous to ATP hydrolysis but without any reliance on biochemistry. 
These two-state nanostructures are functionally similar to a hinge, so that a configurational change mediates the opening and closing of the nanostructure, see Fig.~\ref{fig:dimer_model}\textsf{A}. 
However, to keep our results generally applicable, we do not further specify the nanostructures' geometry or inner workings.
Instead, we show how a ``Source'' structure that is kinetically trapped in a long-lived high-energy state can release this energy after forming a complex with a ``Machine'' structure. The released energy is then harnessed to drive the Machine into its own high-energy state, efficiently and effectively transferring energy between the structures.
This energy-delivery mechanism is achieved through a careful balancing of the energy profiles of the two structures along their reaction coordinates to create a relaxation pathway that has minimal energy barriers and is driven only by the initial conditions. Furthermore, adjusting the energy profiles enhances different aspects of the mechanism, such as the performance, efficiency, and power, making this mechanism highly robust and adaptable, and setting the stage for nontrivial inverse design in a range of experimental settings.

\section{Design principles for controlled energy delivery\label{sec:mechanism}}

\begin{figure*}[t!]
    \includegraphics[width=\textwidth]{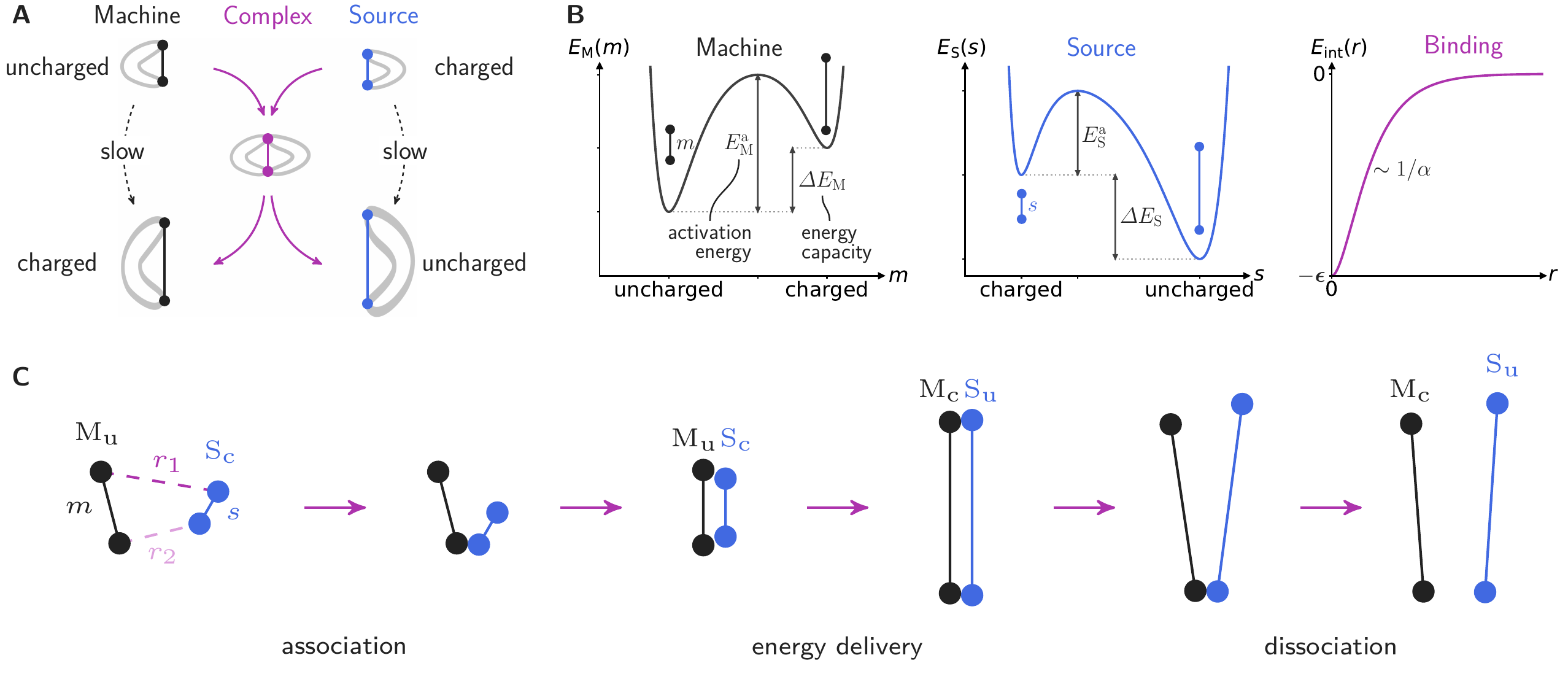}
    \caption{Generic scheme for energy delivery between bistable nanostructures. (\textsf{A}) Two nanostructures, called the ``Machine'' and ``Source'', that can undergo a conformational change that elongates the distance between binding sites at the ends of the structures. We model these structures as dimers, as shown. Although the individual transitions are slow, we show that there is a potentially fast coupled pathway in which the dimers form a complex and transition together (purple). All transitions can in principle go in both directions, but we show the target direction only. (\textsf{B}) The energy of the Machine, $\EM(m)$, and the Source, $\ES(s)$, depend only on the dimer lengths, $m$ and $s$, respectively. As shown in the example energy profiles, each structure has a ``charged'' high-energy state and an ``uncharged'' low-energy state. The Machine and Source dimer ends bind to each other through a Morse interaction energy $E_\text{int}(r)$, with strength $\epsilon$ and range $1/\alpha$. (\textsf{C}) The coupled reaction consists of three parts: association, where two bonds are sequentially formed, energy delivery, where both dimers elongate together, and dissociation, where the bonds are broken. Our goal is to understand how to adjust the energy profiles in \textsf{B} so that this coupled reaction is not only thermodynamically favorable, but fast and efficient enough to effectively power the Machine in some separately designed task.}
    \label{fig:dimer_model}
\end{figure*}
ATP can be viewed very roughly as a two-state molecule whose transition to ADP releases energy, powering various cellular functions. Here, we develop a simple two-state model for nanomachines that is inspired by ATP hydrolysis but not intended to mimic it directly. Rather, we seek a generic platform for studying bistable nanostructures and how their reactions can be coupled using explicit and simple physical interactions.
The key observation is that, regardless of the internal complexity of the nanostructure, its interaction with the outside world is mediated through its binding sites. This means that from the perspective of a second nanostructure, all that matters is where the binding sites are and how much energy is stored in the structure as a function of their separation. We can therefore model our bistable nanostructures as dimers: each is composed of two point-like binding sites whose separation determines the internal energy. The dimer length therefore tracks the separation between the actual binding sites and serves as the reaction coordinate for the conformational change. The internal energy as a function of this length, which we call the energy profile, encodes the full complexity of the underlying conformational transition in a single coarse-grained degree of freedom. This reduction is not an approximation of any specific system, but rather a deliberate abstraction that isolates the minimal physical ingredients required for controlled energy delivery, making our results applicable across a wide range of experimental settings. 

Figure~\ref{fig:dimer_model} outlines the generic setup for energy delivery. 
Both structures, which we call the Machine (black dimer) and Source (blue dimer), have a high-energy, or ``charged'', state, and a low-energy, or ``uncharged'', state (Fig.~\ref{fig:dimer_model}\textsf{A}). Both begin in their shorter state, but with opposite charge designations: the Machine in its uncharged state and the Source in its charged state. This asymmetry is essential so that when the two structures bind, the coupled reaction where they both open will naturally transfer energy from the Source to the Machine.

In isolation, each structure can spontaneously open due to thermal fluctuations, although we will only consider cases where the energy barriers are large, making these reactions slow but still possible. The Machine reaction $\Mu \to \Mc$ is endergonic (thermodynamically unfavorable), while the Source reaction $\Sc \to \Su$ is exergonic (thermodynamically favorable with a net release of free energy). However, the two structures can attach through binding sites at their dimer ends to form a Machine-Source complex, allowing them to transition \emph{together} through the coupled reaction $\Mu + \Sc \to \Mc + \Su$ (Fig.~\ref{fig:dimer_model}\textsf{C}). The complex can then dissociate, leaving the Machine alone in its charged state and Source in its uncharged state, meaning that energy has been transferred. Understanding how to engineer this coupled reaction to be both kinetically fast and thermodynamically favorable is one of the main results of this paper.

Let $m$ and $s$ be the length of the Machine and Source dimers, respectively, and the dimer energies be given by energy profiles $\EM(m)$ and $\ES(s)$, Fig.~\ref{fig:dimer_model}\textsf{B}.
Our model for these energy profiles, discussed in Methods, is intentionally highly parameterized, ensuring two (meta)stable states while providing flexibility in the details, such as the position and energy of the two states. 
The difference in energy between the charged and uncharged states gives the ``energy capacity'' $\varDelta E$ of each structure, and the height of the closed-to-open energy barrier gives the ``activation energy'' $E^\mathrm{a}$. We require the energy capacity of the Source to be greater than that of the Machine, $\varDelta E_\mathrm{S} > \varDelta E_\mathrm{M}$, otherwise energy delivery cannot be thermodynamically favorable. 

The Machine and Source interact through binding sites at the dimer ends through a Morse potential shown in Fig.~\ref{fig:dimer_model}\textsf{B} and given by
\begin{equation}
    E_\mathrm{int}(r) = \epsilon\left(e^{-2\alpha r} - 2e^{-\alpha r}\right),
\end{equation}
where $r$ is the separation between binding sites of different structures.
The binding energy $\epsilon$ will be a key parameter later. 
In principle, there are four possible bonds between the two dimers, but to simplify the analysis, we assume that each dimer end can only bind with a single end of the other dimer, which often can be achieved experimentally through specific interactions. If the relevant binding-site separations are $r_1$ and $r_2$, then the total energy of the system is
\begin{equation}\label{eq:Etot}
    E(m, s, r_1, r_2) = \EM(m) + \ES(s) + E_\mathrm{int}(r_1) + E_\mathrm{int}(r_2).
\end{equation}
In practice, $m$, $s$, $r_1$, and $r_2$ are calculated from the three-dimensional positions of the four dimer ends. 

We will find it useful to consider the minimum energy for fixed dimer lengths $m$ and $s$, \ie, $E_\text{min}(m,s) \equiv \min_{r_1,\,r_2} E(m,s,r_1,r_2)$. Such minima always occur when the two dimers are completely collinear, and we derive analytic expressions for $r_1$ and $r_2$ at the minimum, as well as $E_\text{min}(m,s)$, in Methods. When $m$ and $s$ are close to each other ($\left| m - s \right| \leq \ln 4/\alpha$), the minimum energy state is when $r_1=r_2$, but when $\left| m - s \right| > \ln 4/\alpha$ this symmetric state is unstable and instead one bond dominates (\emph{e.g.}, $r_1\approx 0$). See Methods for the full derivation and analytic form of $E_\text{min}(m,s)$ (Eq.~\eqref{eq:Emin_full}).
We will proceed by studying the energy landscape given by $E_\text{min}(m,s)$ to identify features of the energy profiles that are necessary for the coupled reaction to proceed. While focusing on $E_\text{min}(m,s)$ neglects fast fluctuations of the dimers that keep $m$ and $s$ fixed, all intuition developed by studying $E_\text{min}(m,s)$ will be validated and justified in Section~\ref{sec:MD} via overdamped Langevin Dynamics simulations.

All quantities are given in units of $M$, $\kb$, $T$, and $\sigma$, where $M$ is the mass of the dimer ends, $\kb$ is the Boltzmann constant, $T$ is the temperature, and $\sigma$ is a fundamental unit of length that we set to 1.
Note that rather than fixing $\sigma$ to a characteristic object size, we instead keep $\sigma$ fixed so that dimer lengths are free to vary as design parameters without requiring rescaling of all quantities.
Energies are given in units of $\kT$, time in units of $\sqrt{M \sigma^2 / \kT}$, and the friction coefficient in units of $\sqrt{M \kT / \sigma^2}$.

\subsection{Stability of the Source}
\begin{figure}[htp]
    \includegraphics[width=\linewidth]{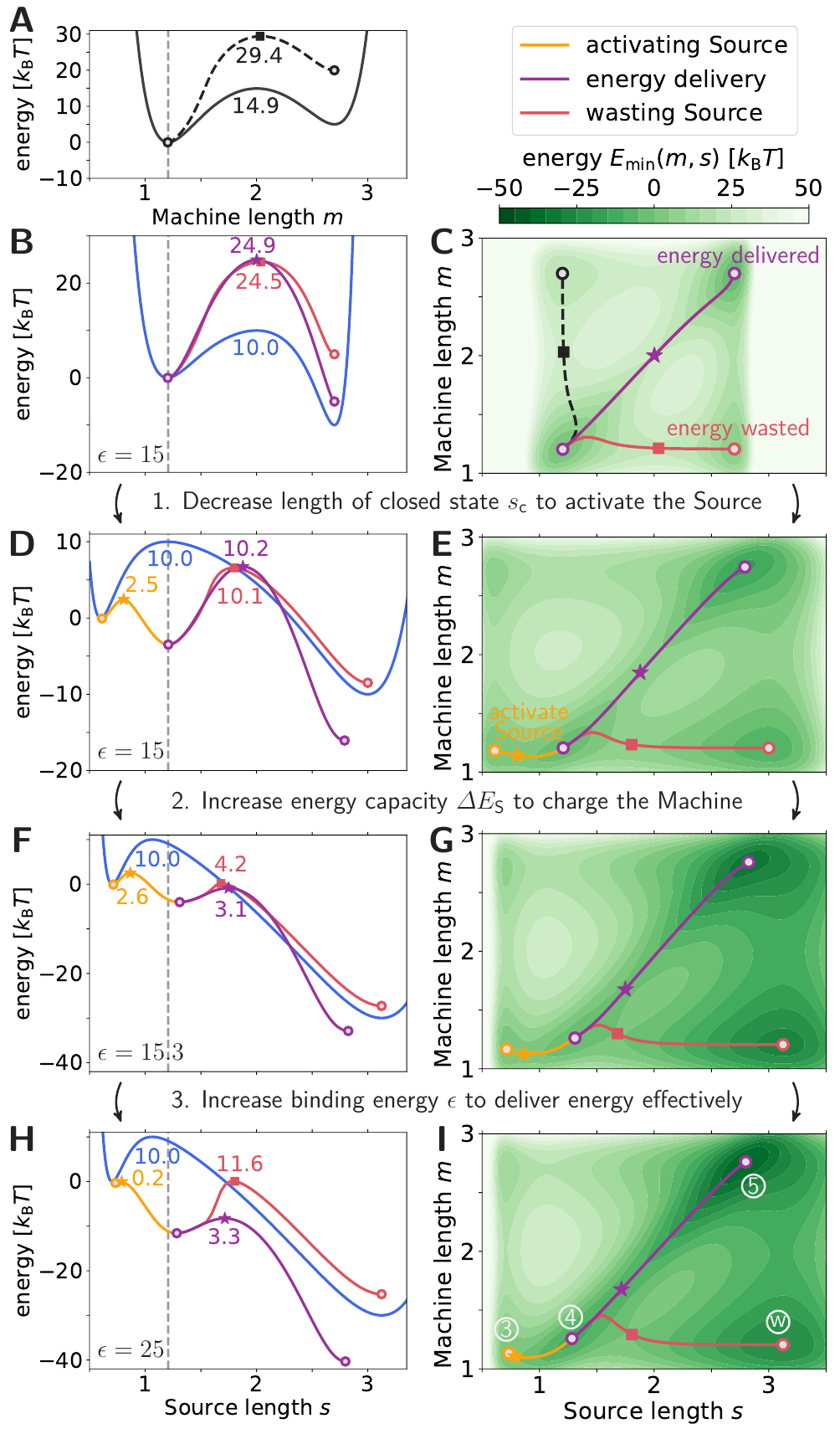}
    \caption{Sculpting a relaxation pathway for controlled energy delivery. The solid black curve in \textsf{A} and blue curves in \textsf{B}, \textsf{D}, \textsf{F}, and \textsf{H} show the energy profiles for the Machine and Source, respectively, discussed in the text. Starting with the generic energy profiles in \textsf{A-B}, we decrease the length of the closed Source state (\textsf{D}), increase the energy capacity of the Source (\textsf{F}), and increase the binding energy $\epsilon$ (\textsf{H}), all with other minor changes as discussed in the text. \textsf{C}, \textsf{E}, \textsf{G}, and \textsf{I} show the corresponding minimum energy $E_\mathrm{min}(m,s)$ for fixed $m$ and $s$. The purple curve shows the desired coupled transition, the orange curve shows the activating transition, the red curve shows the undesired separated transition of the Source being wasted, and the dashed black curve shows the Machine opening on its own, which is always thermodynamically unfavorable but shown in \textsf{A} and \textsf{C} for reference. The transition pathways are obtained using the doubly-nudged elastic band (DNEB) method~\cite{tryg04} and transition states (indicated by stars and squares) are refined using eigenvector following~\cite{cerj81, wale04, maur05}. The colored numbers in the left panels list the forward energy barriers (in units of $\kT$) of the corresponding transitions. The state labels in \textsf{I} refer to Fig.~\ref{fig:md_results}.}
    \label{fig:mechanism_explainer}
\end{figure}

In ATP hydrolysis, ATP is remarkably stable in solution despite the reaction being thermodynamically favorable; without an enzyme, spontaneous hydrolysis is negligibly slow. Our mechanism requires the same of the Source: its charged state must be kinetically protected so that it does not discharge before encountering a Machine. This means that a large activation energy, $E^\mathrm{a}_\mathrm{S}$, is necessary.
However, a large activation energy in the Source presents a problem because when the Source \emph{does} make contact with the Machine, this activation energy must be overcome in order to access the energy in the Source and charge the Machine. 
To see this problem more clearly, consider the generic energy profiles shown in Fig.~\ref{fig:mechanism_explainer}\textsf{A} for the Machine and the blue curve in Fig.~\ref{fig:mechanism_explainer}\textsf{B} for the Source. 

At first, these profiles seem reasonable: the activation energies are large and the Source has a larger energy capacity than the Machine. 
Figure~\ref{fig:mechanism_explainer}\textsf{C} shows $E_\text{min}(m,s)$, the minimum energy of the complex as a function of the two dimer lengths. 
When the complex forms, the system sits in the bottom left minimum, from which there are three possible transitions. 
First, the Machine and the Source can open together, moving to the top right in Fig.~\ref{fig:mechanism_explainer}\textsf{C} (purple curve) -- this is the desired outcome as it drives the Machine to its open, charged state. 
Second, the Machine can stay in its closed, uncharged state while the Source opens, moving to the bottom right in Fig.~\ref{fig:mechanism_explainer}\textsf{C} (red curve) -- this is an undesirable outcome as it wastes the Source without charging the Machine.
Third, the Source can stay in its closed, charged state while the Machine opens, moving to the top left in Fig.~\ref{fig:mechanism_explainer}\textsf{C} (dashed black curve) -- this last reaction is not thermodynamically favorable and can be ignored for now.
Here and in the following, transition pathways between local minima are obtained using the doubly-nudged elastic band (DNEB) method~\cite{tryg04} and transition states are refined using eigenvector following~\cite{cerj81, wale04, maur05}.

Importantly, in this generic example, both the purple and red paths have an even larger energy barrier than the Source's original activation energy $E^\mathrm{a}_\mathrm{S}$, as shown in Fig.~\ref{fig:mechanism_explainer}\textsf{B}. If $E^\mathrm{a}_\mathrm{S}$ is prohibitively large, then so too are these barriers. Therefore, these generic energy profiles will not lead to an energy-delivery reaction.

\subsection{Activating the Source}
In ATP hydrolysis, motor proteins or other ATPase enzymes lower the activation barrier for the ATP-to-ADP transition by binding most strongly to the transition state~\cite{paul48}. For us, the Machine plays the role of the enzyme, using its own geometry to pull the Source towards its transition state and lower the barrier to activation. Following prior work~\cite{agud21,muno23} on mechanical catalysts, this will allow the Source to overcome its energy barrier faster when it is part of the complex.
The key step is to introduce an offset in the length of the dimers in their closed states. 
For example, the blue $\ES(s)$ curve in Fig.~\ref{fig:mechanism_explainer}\textsf{D} and the black $\EM(m)$ curve in Fig.~\ref{fig:mechanism_explainer}\textsf{A} have a clear difference in the length at the closed minima, and the transition state of the Source is roughly aligned with the minimum of the Machine, consistent with Pauling's notion that enzymes bind most strongly to the transition state of a substrate~\cite{paul48}.

Now, as the Source begins to fluctuate open, it is pulled by the Machine due to the attractive interactions between the dimer ends. Thus, part of the energetic cost of opening the Source is compensated by the Machine-Source binding, causing the barrier to drop significantly. This can be seen in the orange curve in Fig.~\ref{fig:mechanism_explainer}\textsf{D}, which has a barrier of only about $2.5\,\kT$, compared to the original activation energy of $10\,\kT$. Importantly, the details of this offset matter, and achieving such a reduced barrier involves a complex interplay between the amount of offset, the binding energy and interaction range, and the curvature of $\ES(s)$ near the top of the barrier. To navigate this complex design space, we have developed a differentiable state-based model that approximates certain dynamical features of the system. This procedure is discussed in detail in the SM, and an illustrative example appears in the next subsection.

Figure~\ref{fig:mechanism_explainer}\textsf{D} shows that activating the Source is possible, resulting in an additional intermediate state that can be reached through a small energy barrier. 
This is progress, as we now have access to the considerable amount of energy stored in the Source. However, Fig.~\ref{fig:mechanism_explainer}\textsf{D} and \textsf{E} also shows that this is not enough, as the purple path still has a prohibitively large barrier.
While the Source now wants to expand, this force is still not enough to charge the Machine.

\subsection{Charging the Machine}

The energy released by ATP hydrolysis does not simply dissipate as heat, but rather is harnessed to drive a conformational change in the motor protein, creating the power stroke that drives function. The analogous step here is using the energy stored in the Source, once activated, to drive the Machine into its own high-energy state. In order to provide the necessary forces, the energy profile of the Source needs to be steeper just to the right of its transition state. Figure~\ref{fig:mechanism_explainer}\textsf{F} accomplishes this while modifying $\ES(s)$ so that the energy capacity is significantly increased, and also the position of the barrier is moved slightly to the left. This has the desired effect as the energy barrier of the purple path is now reduced to $3.1\,\kT$, while only slightly increasing the barrier of the orange path. 
As before, the details of this new energy profile matter, and achieving the right balance between different features (including slopes and curvatures) of $\ES(s)$ is achieved again through our differentiable state-based model.

\begin{figure}[t]
    \includegraphics[width=\linewidth]{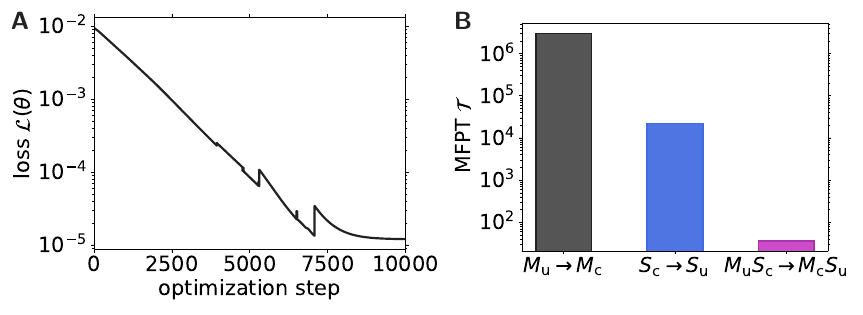}
    \caption{Optimization results of the differentiable state-based model starting from the parameters in Fig.~\ref{fig:mechanism_explainer}\textsf{D-E} and resulting in the parameters in Fig.~\ref{fig:mechanism_explainer}\textsf{F-G}. (\textsf{A}) Loss from Eq.~\eqref{eq:loss_opt} during optimization using the L-BFGS optimizer from JAXopt~\citep{blon22}. Starting from an initial value $\mathcal{L}(\theta_\text{ini})\simeq9.5\cdot10^{-3}$ the loss decreases by almost three orders of magnitude until it converges at an optimized value $\mathcal{L}(\theta_\text{opt})\simeq1.2\cdot10^{-5}$ after $10^4$ optimization steps. The discrete jumps in the loss function occur due to recalculation of the energy landscape every ten optimization steps by recalculating all DNEB pathways. (\textsf{B}) Mean first-passage times of Machine and Source reactions, $\Mu \to \Mc$ and $\Sc \to \Su$, as well as for the coupled reaction $\Mu\Sc \to \Mc\Su$ of the optimized system. The mean first-passage time of the coupled reaction is significantly smaller than the Machine on its own or the Source on its own.}
    \label{fig:optimization}
\end{figure}

We start by using the transition pathways and transition states shown \eg in Fig.~\ref{fig:mechanism_explainer}\textsf{E} to approximate all transition rates~\cite{kram40,hang90}, and calculate the mean first-passage time, $\mathcal{T}_{\Mu\Sc\to\Mc\Su}$ of the coupled reaction, \ie, going from the bottom left state to the top right state. We then construct a loss function 
\begin{equation}\label{eq:loss_opt}
    \mathcal{L}(\theta) = \frac{\mathcal{T}_{\Mu\Sc\to\Mc\Su}(\theta)}{\mathcal{T}_{\Mu\to\Mc}(\theta)} + \mathcal{C}(\theta),
\end{equation}
where we have normalized by the mean first-passage time of the Machine charging on its own. 
We also include constraints $\mathcal{C}(\theta)$ on energy barriers and lengths to ensure that the solutions are physically relevant (see SM).
These quantities depend on the underlying model parameters $\theta$, and we use methods of Automatic Differentiation~\cite{jax2018github,scho21,bayd18,rume86,weng64} to numerically calculate the gradient $\nabla_\theta \mathcal{L}$, allowing us to exploit standard gradient-based optimization routines to minimize $\mathcal{L}$.
Figure~\ref{fig:optimization}\textsf{A} shows $\mathcal{L}$ during optimization, starting from the parameters in Fig.~\ref{fig:mechanism_explainer}\textsf{D-E} and resulting in the parameters in Fig.~\ref{fig:mechanism_explainer}\textsf{F-G}. 
The loss decreases by almost three orders of magnitude, and Fig.~\ref{fig:optimization}\textsf{B} shows that the mean first-passage time of the coupled reaction is significantly smaller than the uncoupled reaction times.
A more detailed discussion can be found in the SM.

However, note that this does not consider the red waste transition, the barrier of which is also significantly reduced, meaning that the Source can easily discharge without charging the Machine. 
We are able to force the system along the desired purple path by making one final modification: we increase the binding energy to $\epsilon=25\,\kT$, Fig.~\ref{fig:mechanism_explainer}\textsf{H} and \textsf{I}.
This makes it harder for the Source and Machine to separate, making it more likely that they transition together, as desired.

\subsection{The dissociation tradeoff}
In the cell, ATP powers motor proteins repeatedly and in large numbers. For example, a single Kinesin hydrolyzes roughly 50 ATP molecules per second~\cite{howa96}.
How can we ensure that our energy-delivery mechanism is not only effective but also repeatable? 
While we just saw that increasing the binding energy prevents the undesired wasteful transition, this binding energy must be overcome for the structures to dissociate. Thus, we are faced with an important tradeoff: strong binding energies are necessary to couple the two reactions, but dissociation is hindered if they are too strong.
This is related to the Sabatier principle, which states that an optimal catalyst should bind to reactants with an intermediate strength to promote the reaction without preventing dissociation~\cite{rivo23}. 
To identify the proper balance between high and low binding energies, we turn, in the next section, to full Brownian Dynamics simulations.
We will see that the mechanism proposed here performs well in this intermediate-binding-energy regime, enabling nontrivial functional behavior.

\section{Verifying and characterizing the mechanism \label{sec:MD}} 

To verify that our mechanism not only works as intended but is also  similarly robust, we now turn to Brownian Dynamics simulations in which an effective bath of Source structures drives a Machine through many repeated charging events, or `actions.'
The structures are modeled as interacting dimers in three dimensions, as described in Section~\ref{sec:mechanism}. 

\begin{figure*}[t!]
    \includegraphics[width=\textwidth]{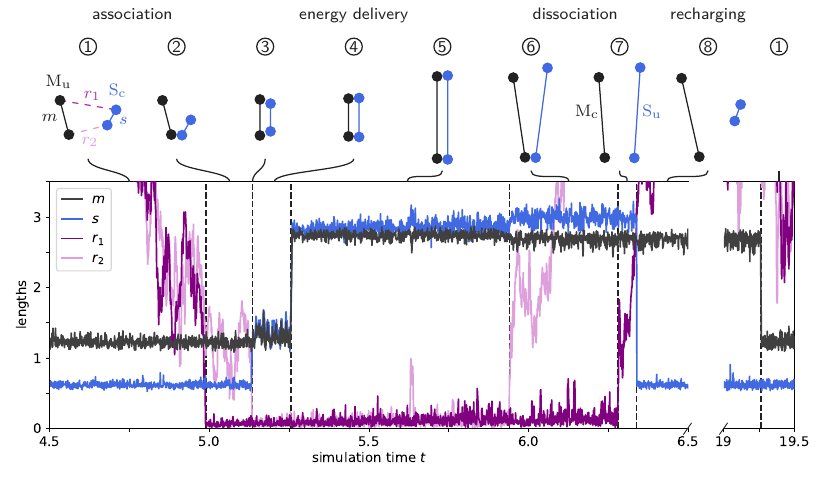}
    \caption{Example trajectory from a Langevin Dynamics simulation demonstrating the progression of the energy-delivery mechanism as discussed in the text. The time evolution of all four relevant lengths, Machine and Source lengths, $m$ and $s$, as well as the dimer-end separations $r_1$ and $r_2$, is shown for a complete Machine action. We use the energy profiles given by $\theta_2$ in Table~\ref{tab:params} and shown in Fig.~\ref{fig:quality}\textsf{G} and \textsf{H} for this representative example. In order to illustrate a complete action over a relatively short time window, this trajectory includes forced dissociation (on a time scale $\tau_\dd=1$), which is introduced and discussed in Section~\ref{sec:forced_dissociation}. Simulations without forced dissociation show the same qualitative progression, see SM for additional trajectories and discussion.}
    \label{fig:md_results}
\end{figure*}

\subsection{An illustrative example}
Figure~\ref{fig:md_results} shows a representative portion of a trajectory that allows us to visualize the progression of the coupled reaction. Shown are the two dimer lengths, $m$ and $s$, and the distances between the attractive ends, $r_1$ and $r_2$, from which we can classify the system into various states to reveal a clear progression through the energy-delivery reaction.
As discussed in Methods, we use a Source-recharging scheme to mimic a bath of Sources so that the energy delivery process can repeat indefinitely.

Initially, the dimers begin in their closed states, $\text{M}_\text{u}$ and $\text{S}_\text{c}$, and are not near each other, \textcircled{\textsf{\scriptsize 1}}. Free to diffuse, the dimers eventually associate, first at one end, \textcircled{\textsf{\scriptsize 2}}, and then at the other, \textcircled{\textsf{\scriptsize 3}}. Once both dimer ends are bound, the Machine applies an outward force on the Source, which almost immediately crosses its barrier, \textcircled{\textsf{\scriptsize 4}}. Note that the lifetime of state \textcircled{\textsf{\scriptsize 3}} is so small that we sometimes have difficulty resolving it, see SM. The Source now applies an outward force on the Machine, helping the Machine to overcome its barrier and bringing both dimers to an open state, \textcircled{\textsf{\scriptsize 5}}. 

Next, the dimers begin to dissociate, first at one end, \textcircled{\textsf{\scriptsize 6}}, and then at the other, \textcircled{\textsf{\scriptsize 7}}. This leaves the Machine in its charged open state, $\text{M}_\text{c}$, and the Source in its uncharged open state, $\text{S}_\text{u}$, meaning we have successfully transferred energy from one structure to the other. Finally, once the Source and Machine are sufficiently separated, the Source is artificially recharged, \textcircled{\textsf{\scriptsize 8}}, to mimic a constant concentration of usable Sources. Independent of this last step, the Machine will eventually and spontaneously transition back to its uncharged state, returning the system to state \textcircled{\textsf{\scriptsize 1}}.
The process shown in Fig.~\ref{fig:md_results} repeats indefinitely -- energy is repeatedly injected through the recharging of the Source, partially delivered to the Machine through the relaxation pathway, and finally dissipated when the Machine transitions back to its uncharged state. 

\begin{figure}[t!]
    \includegraphics[width=\linewidth]{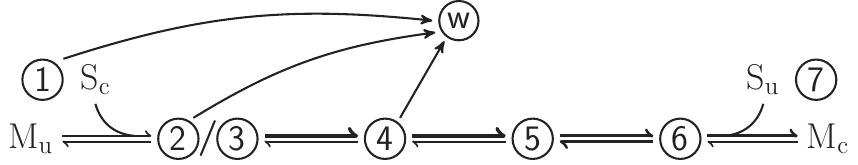}
    \caption{Transition network between meta-stable states in our model. The progression along the horizontal line corresponds to the desired mechanism including the activating and energy-delivery reactions highlighted by thick arrows. Transitions to the waste state are irreversible and undesirable. This network does not consider the recharging of the Source or the spontaneous discharging of the free Machine. The state labels refer to Fig.~\ref{fig:mechanism_explainer}\textsf{I} and Fig.~\ref{fig:md_results}.}
    \label{fig:network}
\end{figure}

This process can be represented as a reaction network, shown in Fig.~\ref{fig:network}. Progression along the horizontal line corresponds to the desired mechanism. Importantly, we include transitions to the waste state, \textcircled{\textsf{\scriptsize w}}, which can be reached by following the red path in Fig.~\ref{fig:mechanism_explainer} or by the Source spontaneously overcoming its activation energy either in state \textcircled{\textsf{\scriptsize 1}} or state \textcircled{\textsf{\scriptsize 2}}. 

Although the reaction network in Fig.~\ref{fig:network} offers a natural framework for optimization, much like the smaller-scale state-based model discussed above, there is a fundamental tension with this approach. The energy-delivery mechanism succeeds because the coupled reaction proceeds through very small energy barriers on the order of $\kT$ where the rate-limiting steps are accelerated by the Machine-Source coupling. However, these small energy barriers are precisely where Kramers theory and other state-based estimates for transition rates become highly inaccurate, with errors that compound over the entire reaction network (see SM). Attempting to optimize the full dynamics via the reaction network would thus mean fine-tuning parameters within the uncertainty range of the model itself, which is a futile exercise. Instead, we use our state-based model as a guide to identify promising energy profiles and parameter regimes, and then rely on Langevin Dynamics simulations to evaluate these designs. This approach trades analytic optimization for numerical fidelity, allowing us to demonstrate and characterize the principles underlying energy delivery while respecting the physics of the system.

\subsection{Numerical characterization}
To characterize our mechanism, we define three measures of merit. First, the performance, $p$, quantifies the probability of the Machine being in a usable charged state normalized by the equilibrium probability without a Source structure.\footnote{A ``usable'' charged state is one where the Machine is free to discharge, which excludes the doubly-bound state \textcircled{\textsf{\tiny 5}}. In the SM, we also consider a stricter definition that requires the Source to be completely detached.} A value $p>1$ means that the Machine is being actively driven to its charged state and the mechanism is working as intended, while $p<1$ means that the presence of the Source actually hurts the Machine by trapping it in the complex. 
Next, the power, $\mathcal{P}$, quantifies the rate at which work can be done by the Machine. It is based on the increase in Machine charging events -- which we call Machine actions -- in the presence of the Source compared to the Machine alone. Finally, the efficiency, $\eta$, quantifies the fraction of stored energy in a Source structure that is successfully delivered to a Machine. This combines two sources of energy loss: energy dissipated during an exchange process as well as when a Source spontaneously discharges without transferring any energy to the Machine.
The relative importance of these quantities depends on the application, and in principle one can find energy profiles that prioritize certain quantities over others.
Exact definitions of these measures appear in Methods.

\begin{figure*}[t!]
    \includegraphics[width=\textwidth]{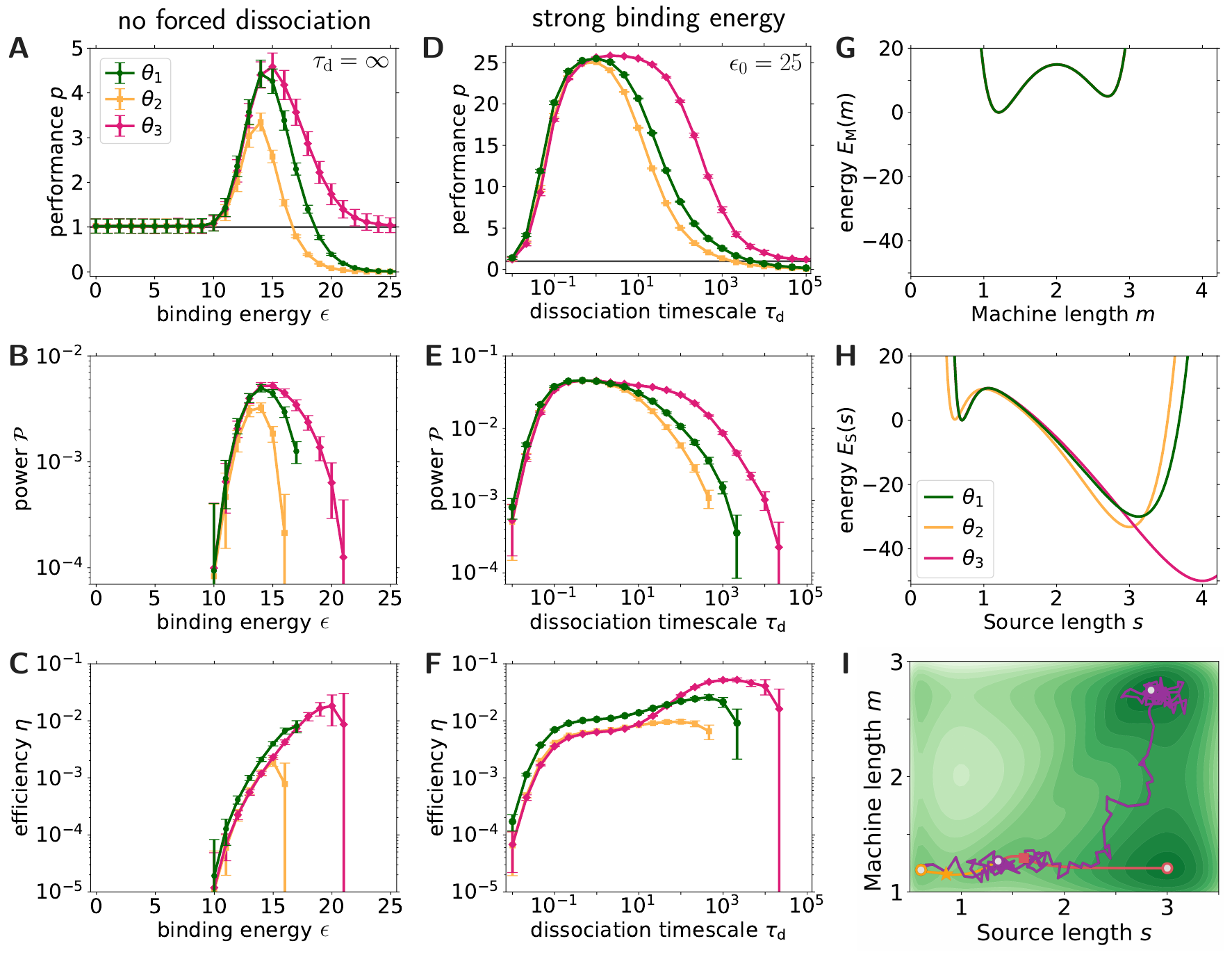}
    \caption{Measures of merit of the energy-delivery mechanism for three designs with different energy profiles. (\textsf{A-C}) Performance $p$, power $\mathcal{P}$, and efficiency $\eta$ as functions of the binding energy $\epsilon$ without forced dissociation ($\tau_\dd=\infty$). The design with parameters $\theta_1$ reaches its maximum values $p_{\max}(\epsilon^*) = 4.43 \pm 0.31$ and $\mathcal{P}_{\max}(\epsilon^*) = (4.97 \pm 0.40)\cdot10^{-3}$ at $\epsilon^* = 14$ with an efficiency of $\eta(\epsilon^*) = (2.10 \pm 0.17)\cdot10^{-3}$, and corresponds to the optimized parameters shown in Figs.~\ref{fig:mechanism_explainer}\textsf{F-I} and \ref{fig:optimization}. The second design performs slightly worse with maximum values $p_{\max}(\epsilon^*) = 3.31 \pm 0.20$ and $\mathcal{P}_{\max}(\epsilon^*) = (3.25 \pm 0.34)\cdot10^{-3}$ with $\eta(\epsilon^*) = (1.23 \pm 0.13)\cdot10^{-3}$ at $\epsilon^* = 14$. The third design performs well over a larger region and shows the highest performance $p_{\max}(\epsilon^*) = 4.58 \pm 0.31$ and power $\mathcal{P}_{\max}(\epsilon^*) = (5.24 \pm 0.38)\cdot10^{-3}$ at $\epsilon^* = 15$, but has the lowest efficiency with $\eta = (1.18 \pm 0.09)\cdot10^{-3}$ at $\epsilon = 14$. Interestingly, large binding energies exhibit increasing efficiency by suppressing the transition to the waste state, with peak efficiencies $\eta_{\max} = (8.09 \pm 1.94)\cdot10^{-3}$ at $\epsilon = 17$ ($\theta_1$), $\eta_{\max} = (1.93 \pm 0.34)\cdot10^{-3}$ at $\epsilon = 15$ ($\theta_2$), and $\eta_{\max} = (1.83 \pm 1.02)\cdot10^{-2}$ at $\epsilon = 20$ ($\theta_3$). We report mean values of 200 long simulations with standard deviation of the mean (error bars). (\textsf{D-F}) Performance, power, and efficiency as functions of the dissociation timescale $\tau_\dd$ using a strong initial binding energy of $\epsilon_0=25\,\kT$. The design with parameters $\theta_1$ reaches its maximum performance $p_{\max}(\tau_\dd^*) = 25.50 \pm 0.04$ at $\tau_\dd^* = 1$ with a power of $\mathcal{P}(\tau_\dd^*) = (4.47 \pm 0.12)\cdot10^{-2}$ and an efficiency of $\eta(\tau_\dd^*) = (1.06 \pm 0.28)\cdot10^{-2}$. The corresponding values for the other designs are $p_{\max}(\tau_\dd^*) = 25.07 \pm 0.05$, $\mathcal{P}(\tau_\dd^*) = (4.45 \pm 0.13)\cdot10^{-2}$, $\eta(\tau_\dd^*) = (6.57 \pm 0.19)\cdot10^{-3}$ at $\tau_\dd^* = 1$ ($\theta_2$), and $p_{\max}(\tau_\dd^*) = 25.85 \pm 0.04$, $\mathcal{P}(\tau_\dd^*) = (4.32 \pm 0.11)\cdot10^{-2}$, $\eta(\tau_\dd^*) = (6.57 \pm 0.16)\cdot10^{-3}$ at $\tau_\dd^* \simeq 2.15$ ($\theta_3$). We report mean values of ten long simulations with standard deviation of the mean. The horizontal line in \textsf{A} and \textsf{D} indicates a performance of $p=1$. We exclude negative values of power and efficiency in \textsf{B-C} and \textsf{E-F} when the number of Machine actions in the presence of the Source drops below the value for the Machine alone (see Methods and SM). (\textsf{G-H}) Energy profiles corresponding to the presented data with parameters listed in Table~\ref{tab:params}. The designs presented here share the same Machine profile and differ in the Source profile. The behavior of additional designs with varying Machine and Source profiles is discussed in the SM. (\textsf{I}) Example Brownian Dynamics trajectory connecting the closed-closed to the open-open complex in a system with parameters $\theta_2$ and constant binding energy $\epsilon=10\,\kT$. The color scale is identical to Fig.~\ref{fig:mechanism_explainer}.}
    \label{fig:quality}
\end{figure*}

Figure~\ref{fig:quality}\textsf{A-C} uses these measures to demonstrate the robustness of our energy-delivery mechanism over a broad range of binding energies for three example energy profiles. Figure~\ref{fig:quality}\textsf{A} shows $p$ as a function of the binding energy $\epsilon$. As expected, we observe a non-monotonic behavior with optimal performances at intermediate binding energies. This result resembles the Sabatier principle in heterogeneous catalysis, where an optimal catalytic performance is achieved at intermediate binding energies, reflecting a symmetry between binding and release~\cite{rivo23}. 
The design with parameters $\theta_1$ corresponds to the optimized parameters shown in Figs.~\ref{fig:mechanism_explainer}\textsf{F-I}, and has a peak performance of $p=4.43 \pm 0.31$ at $\epsilon=14\,\kT$. 
The design with parameters $\theta_2$ was obtained in another optimization scheme similar to the one presented above but starting from different initial parameters. It generally performs slightly worse, with a peak performance of $p=3.31 \pm 0.20$. 
The design with parameters $\theta_3$ is a modification of the first design, where the Source energy capacity and location of the uncharged Source state are manually increased while keeping the slope just to the right of the transition state roughly unchanged. These modifications are intended to promote the dissociation of one of the bonds, allowing for slightly larger binding energies. Indeed, this design exhibits the largest performance of $p=4.58 \pm 0.31$ at $\epsilon=15\,\kT$.

In all cases, the performance approaches 1 at low binding energies because the two structures do not sufficiently interact.  
At high binding energies, the performance of the first two designs drops to $p=0$ because, without the ability to dissociate, the system gets stuck in the fully-bound state \textcircled{\textsf{\scriptsize 5}}. Interestingly, the third design plateaus instead at $p=1$. This is because the difference in the length of the open states causes one bond to break naturally even for high $\epsilon$. Then, while the second bond is effectively permanent, the Machine is nevertheless free to open and close as if the Source was not present at all. 

Figure~\ref{fig:quality}\textsf{B-C} shows $\mathcal{P}$ and $\eta$ for the same simulations. Interestingly, as the binding energy decreases below $\epsilon \approx 15 k_BT$, the performance, power, and efficiency all decrease: there is no clear advantage to having small binding energies. However, the same is not true for large binding energies. While the power decreases qualitatively with performance, large binding energies exhibit increasing efficiency by suppressing the transition to the waste state. Therefore, for experimental objectives that prioritize efficiency, slightly larger $\epsilon$ may be preferable despite the decrease in performance.

\subsection{Forced dissociation}\label{sec:forced_dissociation}

The above results robustly demonstrate energy delivery for intermediate binding energies. Provided a Source structure is prepared in its high-energy state, energy delivery to the Machine is the natural result of passive relaxation. Increasing the binding energy too much hurts this process by increasing the dissociation time, making it harder for subsequent Source structures to deliver their energy. However, if one could force dissociation on faster time scales in the high-binding-energy regime, the mechanism has considerable potential for dramatically increasing the efficiency and performance.

In practice, there are numerous such approaches for forcing dissociation, and facilitated dissociation occurs naturally, for example in the dissociation of transcription factors from DNA~\cite{kama17}. One could increase the energy capacity of the Source and use this to force dissociation, for example through a mechanical process that blocks the binding site as the Source approaches its uncharged state. Alternatively, one could employ a vitrimer-like bond-swap mechanism~\cite{deni16,scio17}, where the Source-Machine bond is exchanged for a bond between the Source and a separate structure, all at constant energy. Finally, one could directly design bonds that weaken over time~\cite{sahu09} or, if experimentally acceptable, force dissociation through global fields or temperature changes.

To study the effect of these approaches generically, we now consider binding energies that decrease with time, so that $\epsilon(t) = \epsilon_0\,e^{-(t-t_0)/\tau_\dd}$, where $t_0$ is the time at which the bond forms. $\tau_\dd$ sets an upper bound for the dissociation time scale, and we can smoothly approach the ``normal'' case of constant binding energy by taking $\tau_\dd \to \infty$. Importantly, this weakening of $\epsilon$ breaks time-reversal symmetry and effectively injects energy into the system, meaning it should not be used on its own to validate our mechanism. Rather, this gives insight into the potential for increased performance under a range of experimental approaches that force dissociation on a time scale $\tau_\dd$.

Figure~\ref{fig:quality}\textsf{D-F} shows the three measures of merit over a broad range of $\tau_\dd$ for the same three designs but with a large binding energy of $\epsilon_0=25\,\kT$.\footnote{Note that for finite $\tau_\dd$, the efficiency shown in Fig.~\ref{fig:quality}\textsf{F} does not consider the energetic cost of forced dissociation.} We observe non-monotonic behavior with an optimum at intermediate dissociation timescales for all designs. The three designs all have peak performances around $p\approx 25$ and for dissociation time scales between $\tau_\dd = 1$ and $\tau_\dd = 10$. This optimal range for forced dissociation can be understood from the time it takes to transition from state \textcircled{\textsf{\scriptsize 1}} to state \textcircled{\textsf{\scriptsize 5}}. As shown in the SM, this is roughly $\mathcal{T}_{1\to5} \approx 4.6$ for $\epsilon=25$, suggesting that a comparable $\tau_\dd$ should maximize performance. Interestingly, for $\tau_\dd=\infty$, the spontaneous dissociation time roughly matches $\mathcal{T}_{1\to5}$ when $\epsilon \approx 14$ (SM), matching the binding energy where we observe peak performance in Fig.~\ref{fig:quality}\textsf{A} and giving insight into the importance of dissociation in our mechanism. Together, these results clearly elucidate the nontrivial tradeoffs when changing the energy profiles, $\epsilon$, or $\tau_\dd$. See the SM for more data describing the behavior of additional designs.

\section{Discussion}

We have presented and numerically validated a robust mechanism for transferring energy between two-state nanostructures that is analogous to ATP hydrolysis. 
This mechanism solves a key hurdle in the pursuit of ensembles of synthetic, powered nanomachines that perform tasks on par with, for example, motor proteins. 
Like ATP, the Source structure begins in a metastable high-energy state, releasing its energy when triggered to do so by the Machine structure. 
Unlike hydrolysis, however, energy delivery is achieved exclusively through physical, rather than biochemical, interactions, meaning that our mechanism is applicable in a wide variety of experimental settings.
One of the major findings of this study is that energy delivery occurs through a two-stage coupled relaxation process: the Machine first lowers the effective barrier for Source activation, and the activated Source then drives the Machine into its charged state.
To achieve this, we have shown that the energy profiles of the two structures can be adjusted to reveal a low-energy relaxation pathway that allows them to undergo conformational transitions together as a complex. 

\begin{figure}
    \includegraphics[width=0.9\linewidth]{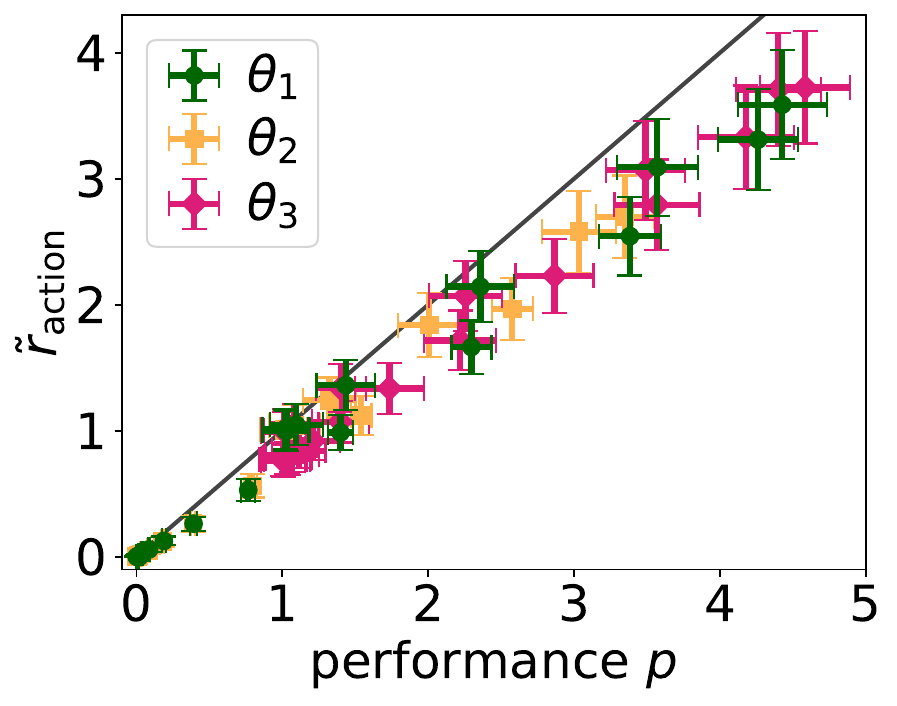}\centering
    \caption{Comparison of the performance and the relative action rate $\tilde r_\mathrm{action}$, defined in Eq.~\eqref{eq:relative_action_rate}, for the data presented in Fig.~\ref{fig:quality}\textsf{A-C}. The performance is slightly larger than the relative action rate leading to the data deviating slightly from the straight line indicating equality. Absolute values of the numbers of Machine actions are shown in the SM.}
    \label{fig:actions}
\end{figure}

After this coupled transition, the Source and Machine structures must dissociate so that the Machine can use this newfound energy and the process can repeat with a new Source. 
We have shown that this can be achieved through intermediate binding energies where we achieve a maximum performance of around $p\simeq 4.6$.
The performance measures the increase in the likelihood of the Machine being in its charged state, excluding times when it is fully bound to the Source (state \textcircled{\textsf{\scriptsize 5}} in Fig.~\ref{fig:md_results}), caused by the presence of the Source. Figure~\ref{fig:actions} shows that the performance gives a good estimate for the relative action rate, which is defined in Eq.~\eqref{eq:relative_action_rate} in Methods. This measures the number of actions (uncharged-to-charged-to-uncharged transitions) of the Machine per unit time in the presence of the Source divided by the number of actions without the Source. Figure~\ref{fig:actions} shows a peak relative action rate nearing 4.
This relationship makes intuitive sense since the rate of discharging is not strongly affected by the presence of the Source.

This connection between performance and the relative action rate allows us to speculate on what could be achieved with the performances we observe.
For example, for the task of maintaining a steady-state concentration difference of a molecule across a barrier, a Machine designed to be a synthetic pump could maintain a roughly $4$ times larger concentration of the molecule on one side of the barrier compared to the other. Alternatively, for the task of walking unidirectionally along a track, a synthetic walker could take roughly $4$ steps to the right for every 1 step it takes to the left. These examples, which are detailed extensively in the SM, demonstrate nontrivial out-of-equilibrium behavior that could provide powerful quantitative solutions to real problems. 
However, note that the efficiencies of the present designs remain modest, reflecting the fact that many Source discharging events are not properly harnessed to charge the Machine. Therefore, while our results demonstrate controlled, repeatable energy delivery, the optimization of efficiency is left as a central target for future work. 

We have identified robust design principles that lead to good performance, most notably the offset in the lengths between binding sites in the closed states, related to the Pauling principle~\cite{paul48}, and the necessity of balancing the forces between the structures with the energy profiles, \eg see Fig.~\ref{fig:mechanism_explainer}. 
To find and study these design principles, we developed a differentiable state-based model that identifies local minima and transition states in $E_\text{min}(m,s)$ to predict the dynamics of the complex. However, while this model is useful, it is insufficient for proper optimization of the full reaction due to approximations and assumptions inherent in transition state theory and because it does not include association/dissociation. We do not claim to have found \emph{optimal} energy profiles, and expect significantly more performant profiles could be found. 

Interestingly, Fig.~\ref{fig:quality}\textsf{I} shows that for the system with parameters $\theta_2$ (orange data in Fig.~\ref{fig:quality}) at $\epsilon=10\,\kT$, there is no transition state that directly connects the closed-closed complex to the open-open complex, similar to the purple pathway connecting states \textcircled{\textsf{\scriptsize 4}} and \textcircled{\textsf{\scriptsize 5}} in Fig.~\ref{fig:mechanism_explainer}\textsf{I}. Nevertheless, Fig.~\ref{fig:quality}\textsf{I} shows that individual trajectories can still connect these states by traversing along contour lines, which allows the mechanism to be effective. This demonstrates the perils of the state-based model, which is not equipped to predict such trajectories. Therefore, we expect that trajectory-based optimization approaches, such as those in Refs.~\cite{bolh23,han25}, are required to optimize the behavior in this regime. 

While our mechanism focuses on two-state nanostructures, each with high- and low-energy states, other paradigms for energy delivery exist. 
For example, one could replace our Machine nanostructure with two rigid structures that one wishes to actively dissociate. This could be achieved with our mechanism, and the similarity between the dissociation of a dimer and the elongation of a two-state dimer is discussed in Ref.~\cite{chat25}.
Furthermore, Albaugh and Gingrich~\cite{alba22} developed a model molecular motor where a particle exits a cage, lowering the total free energy while biasing a cyclic reaction. 

Our mechanism is also complementary to recent work by Chatzittofi \etal~\cite{chat25}, who demonstrated a momentum-based mechanism for fueled enzymatic behavior. The key distinction is that our mechanism inhibits unwanted energy dissipation in the Source by kinetically protecting the meta-stable charged state, whereas the fueled enzyme in Ref.~\cite{chat25} undergoes a continuous downhill cycle whether or not it is attached to a substrate molecule. The price we pay for this efficiency and control is more restrictive requirements on the energy profiles. 
Furthermore, the energy profiles we consider are explicit functions of the structures' conformation (dimer length), meaning we can fully describe the coupled reaction without implicit internal variables, as in Ref.~\cite{chat25}. 
Thus, while Chatzittofi \etal describe a catalyzed coupled reaction, $\text{E} + \Sc + \Mu \to \text{E} + \Su + \Mc$, with an implicit description of the ``Source'' and a kinetic-based mechanism, we provide a fully explicit description of a spontaneous coupled reaction, $\Sc + \Mu \to \Su + \Mc$ that can function in the overdamped regime. Together, these results form a powerful foundation for the \emph{de novo} design of controlled energy transduction.

By demonstrating ATP-hydrolysis-like functionality through purely physical forces, this work suggests a route toward synthetic nanostructures that exchange energy in a controlled and reusable way, without relying on biochemistry. While our results present a proof-of-principle demonstration, further work will be needed to understand how this mechanism performs in more realistic settings. As discussed above, proper optimization will likely require trajectory-based approaches rather than the state-based model developed here, particularly in regimes where successful pathways are not well described by steepest descent paths. Beyond this, our coarse-grained dimer representation was chosen to isolate the minimal physical ingredients required for controlled energy delivery, but implementation-specific models will be needed to assess particular experimental platforms. Such models should include the full geometry of the nanostructures and a validated microscopic interaction model. Incorporating such details will enable the mechanistic understanding developed here to be translated into quantitative design rules for specific synthetic nanomachines.

Achieving optimal energy profiles required by our mechanism remains an experimental challenge, but several existing platforms already provide many of the necessary ingredients. 
In DNA-based nanotechnology, the energy landscape of a bistable structure can be rationally tuned through strand length and sequence. Shape-shifting colloidal assemblies~\cite{good21} offer a complementary route, where designer interactions enable a cluster to undergo collective conformational transitions. Finally, \emph{de novo} protein design~\cite{prae23} represents perhaps the most powerful long-term avenue, as modern computational tools now allow the design of proteins with prescribed energy minima. In all these cases, the primary challenge will be designing the necessary energy profiles of the structures. 
A promising and general design paradigm, inspired by natural bistable proteins~\cite{pdb101,peru78,fisc11}, is to separate the structure into a ``hinge'' region that controls the energetics of the conformational transitions and rigid ``arm'' domains that control the geometry and coupling between structures. Such a hinge-arm architecture could allow the details of the energy profiles to be tuned independently, directly addressing a key design challenge identified in this work.

\section{Materials and methods}

\subsection{Model details}

We model the energy profiles of the Machine and Source through double-well potentials, allowing us to independently tune the positions of the closed, open, and transition states, all barrier heights, and the steepness of the barriers. Each energy profile is a continuous and differentiable combination of two Morse potentials
\begin{align}
    E(r) &= \epsilon_\uu \left(e^{-2\alpha_\uu (r - r_\uu)} - 2e^{-\alpha_\uu (r - r_\uu)}\right) + cr \nonumber\\ 
         &\quad + \epsilon_\cc \left(e^{-2\alpha_\cc (r_\cc - r)} - 2e^{-\alpha_\cc (r_\cc - r)}\right), \label{eq:double_well_potential}
\end{align}
with seven parameters. In Eq.~\eqref{eq:double_well_potential}, $r$ is the dimer length, which is either $m$ or $s$ in the main text depending on whether it is the Machine or the Source.
The purely attractive interaction between the dimer ends of Machine and Source at distance $r$ is modeled by a Morse potential
\begin{equation}
    E_\text{int}(r) = \epsilon\left(e^{-2\alpha r} - 2e^{-\alpha r}\right), 
\end{equation}
with binding energy $\epsilon$ and interaction range $1/\alpha$. There are four potential bonds between the beads of the two dimers to take into account if the dimer ends are symmetric, although the physics of the system is not changed if one of the dimers is inverted. For the sake of clarity in the analysis, we treat the dimer ends as nonsymmetric, and design the interactions such that each dimer end can only bind with a single end of the other dimer, with binding-site separations $r_1$ and $r_2$. Importantly, this choice has no effect on any of our results beyond a marginally longer initial association time, but this is never the limiting step. Altogether, the full energy of the system as function of dimer lengths $m,s$ and separations $r_1,r_2$ is given by Eq.~\eqref{eq:Etot} with a total of 16 parameters $\theta$. 
The physical model parameters $\theta$ used for the results in Figs.~\ref{fig:mechanism_explainer} to \ref{fig:quality} are listed in Table~\ref{tab:params}.
\begin{table*}[t!]\centering
  \setlength{\tabcolsep}{3pt}
  \begin{tabular}{lllllllllllllllll}
   \toprule
    & \multicolumn{7}{l}{Machine} & \multicolumn{7}{l}{Source} & \multicolumn{2}{l}{Binding} \\
    & $m_\uu$ & $m_\cc$ & $\epsilon_\uu^\text{M}$ & $\epsilon_\cc^\text{M}$ & $\alpha_\uu^\text{M}$ & $\alpha_\cc^\text{M}$ & $c^\text{M}$ & $s_\cc$ & $s_\uu$ & $\epsilon_\cc^\text{S}$ & $\epsilon_\uu^\text{S}$ & $\alpha_\cc^\text{S}$ & $\alpha_\uu^\text{S}$ & $c^\text{S}$ & $\epsilon$ & $\alpha$ \\ 
    \midrule
    $\mathcal{\theta}_1$ & $1.2$ & $2.71$ & $26.04$ & $20.72$ & $2.65$ & $2.63$ & $-0.10$ & $0.69$ & $3.01$ & $17.90$ & $28.39$ & $7.34$ & $1.12$ & $-9.96$ & $15.26$ & $5$ \\
    $\mathcal{\theta}_2$ & $1.2$ & $2.71$ & $26.04$ & $20.72$ & $2.65$ & $2.63$ & $-0.10$ & $0.60$ & $3$ & $15.61$ & $57.08$ & $6.83$ & $1.08$ & $-0.06$ & $25$ & $5$ \\
    $\mathcal{\theta}_3$ & $1.2$ & $2.71$ & $26.04$ & $20.72$ & $2.65$ & $2.63$ & $-0.10$ & $0.69$ & $4.00$ & $16.85$ & $86.66$ & $7.64$ & $0.63$ & $-0.19$ & $25$ & $5$ \\
   \bottomrule
  \end{tabular}
  \caption{List of all model parameters in different designs. The parameters $\theta_1$ are used for the main results in Figs.~\ref{fig:mechanism_explainer}, \ref{fig:optimization}, and \ref{fig:quality} and the parameters $\theta_2$ are used in Fig.~\ref{fig:md_results}. While the three designs all follow our design principles, $\theta_3$ is a modification of $\theta_1$, where the Source energy capacity and location of the uncharged Source state are increased while keeping the slope just to the right of the transition state roughly unchanged. The Source profile in design $\theta_2$ has lowered locations of minima and transition state and slightly modified slopes. The measures of merit of the energy-delivery mechanism for all three different designs are analyzed in Fig.~\ref{fig:quality}, where the binding energy $\epsilon$ is varied in \textsf{A-C}.}
  \label{tab:params}
\end{table*}

For fixed dimer lengths $m$ and $s$, we define the minimum energy of the system as
\begin{equation}
    E_\text{min}(m,s) \equiv \min_{r_1,\,r_2} E(m,s,r_1,r_2),
\end{equation}
where $E(m,s,r_1,r_2)$ is given by Eq.~\eqref{eq:Etot}. 
This minimum always occurs when the dimers are collinear, with the ends of the shorter dimer located at or within the longer dimer. This means that
\begin{equation}
    |m - s| = r_1 + r_2.
\end{equation}
We therefore can set $r_1=|m-s|-r_2$ in Eq.~\eqref{eq:Etot} and find stable configurations by setting to zero the derivative of $E$ with respect to $r_2$.
First, if $|m-s| \leq \ln4/\alpha$, then the energy is minimized when both potential bonds are active. However, due to the difference in $m$ and $s$, the separations $r_1$ and $r_2$ cannot be zero, but are instead 
\begin{align}
    r_1 = r_2 = \frac{|m-s|}{2}.
\end{align}
However, if $|m-s| > \ln4/\alpha$, then one of the bonds will dominate, although the other bond still pulls slightly. Without loss of generality, we can assume that the dominant bond is $r_2$, in which case we have
\begin{align}
    r_1 = |m-s| - \ln\chi/\alpha, && r_2 = \ln\chi/\alpha,
\end{align}
where
\begin{equation}
    \chi \equiv \frac{1}{2}\left(e^{\alpha|m-s|} - e^{\alpha|m-s|/2} \sqrt{e^{\alpha|m-s|}-4}\right).
\end{equation}
Note that $\chi$ is strictly real and positive when $|m-s| > \ln4/\alpha$.
The minimum energy thus reads
\begin{widetext}
\begin{equation}
    E_\mathrm{min}(m,s) = E_\text{M}(m) + E_\text{S}(s) + 
        \begin{cases}
            2\epsilon\left(e^{-\alpha|m-s|} - 2e^{-\alpha |m-s|/2}\right), & \text{if } |m-s| \leq \ln4/\alpha,\\
            \epsilon\left(\chi^2e^{-2\alpha|m-s|} - 2\chi e^{-\alpha|m-s|} - \frac{2}{\chi} + \frac{1}{\chi^2}\right),  & \text{if } |m-s| > \ln4/\alpha.
        \end{cases}
        \label{eq:Emin_full}
\end{equation}
\end{widetext}

\subsection{Overdamped Langevin Dynamics simulations}

We run overdamped Langevin Dynamics simulations with friction coefficient $\gamma=0.1$ and thermal energy $\kT=1$ using a time step $\dt=10^{-5}$ for $2\cdot10^{10}$ steps such that we observe multiple Machine actions (see Sec.~\ref{sec:mechanism} for discussion of units). We simulate the dimers in a three-dimensional box of side length 10 with periodic boundary conditions, starting from a random configuration initialized with the dimers separated and in their closed state. 
For each data point in Figs.~\ref{fig:quality} and \ref{fig:actions}, we run either 200 (Fig.~\ref{fig:quality}\textsf{A-C}) or ten (Fig.~\ref{fig:quality}\textsf{D-F}) such simulations with different seeds and analyze them over time windows $T_\text{sim}$ from the first to the last Machine charging event. This removes bias caused by the initial conditions.

In order to replicate the effect of a constant concentration of usable charged Source structures, we introduce a recharging scheme that converts an uncharged Source structure into a charged Source structure. Recharging only occurs when the Source and Machine are sufficiently separated  (see SM). Note that while this directly injects energy into the system, the coupled energy-delivery reaction itself is a purely dissipative process that is driven only by the initial conditions. For simulations with finite $\tau_\dd$, recharging the Source resets the binding strength $\epsilon$ to its initial $\epsilon_0$.

\subsection{Measures of merit}

We define the performance
\begin{equation}
    p \equiv P_\text{usable}^\text{with Source} \; / \; P^\text{Machine only}_\text{usable},
\end{equation}
to be the probability of the Machine being in a usable charged state in the presence of the Source normalized by the same probability with only the Machine. $P_\text{usable}$ is calculated from the fraction of time where the Machine length $m$ is larger than the length at the energy barrier and the Machine-Source is not in a doubly-bound complex, which indicates how often the Machine is ready to use this energy to perform a task. For the Machine only, \ie, when no Source is present, this can be calculated analytically to obtain $P^\text{Machine only}_\text{usable}$ from the Boltzmann equilibrium probability distribution (see SM). The performance $p$ is their ratio.

The power
\begin{equation}
    \mathcal{P} \equiv \frac{\varDelta N_\text{actions}}{T_\text{sim}} \varDelta E_\text{M}
\end{equation}
quantifies the rate that work can be done by the Machine beyond what can be done without the Source. The relevant quantity is the number of charging events (uncharged-to-charged-to-uncharged transitions) observed in the Machine in a time window $T_\text{sim}$, which we call the number of Machine actions. The power is calculated from the excess number of Machine actions in the presence of the Source compared to the equilibrium system of the Machine alone, 
\begin{equation}
    \varDelta N_\text{actions} \equiv N_\text{actions}^\text{with Source} - N_\text{actions}^\text{Machine only},
\end{equation}
times the energy capacity of the Machine. 
We define the relative action rate as their ratio
\begin{equation}
    \tilde{r}_\mathrm{action} \equiv N_\text{actions}^\text{with Source} \; / \; N_\text{actions}^\text{Machine only}, \label{eq:relative_action_rate}
\end{equation}
measuring the number of actions of the Machine per unit time in the presence of the Source divided by the number of actions without the Source.
The power is qualitatively similar to the performance given the relation between actions and performance (see Fig.~\ref{fig:actions}), and is positive roughly when $p>1$ (see SM). 

The energy efficiency $\eta_\text{e} \equiv \varDelta\EM / \varDelta\ES$ of a single energy-transfer event is the ratio of the energy capacities. When $\eta_\text{e}=1$, $100\,\%$ of the Source's energy is transferred to the Machine, and $\eta_\text{e} < 1$ is required for the mechanism to be thermodynamically favorable. Additionally, the number efficiency $\eta_\text{n} \equiv \varDelta N_\text{actions} / N_\text{re}$ is the ratio of the number of additional Machine charging events (compared to equilibrium) to the number of Source recharging events. When $\eta_\text{n}=1$, then every charged Source eventually charges the Machine, providing an action that would not have occurred in equilibrium. However, transitions to the waste state and spontaneous Source-discharging events cause this to be lower. Notice also that this is necessarily concentration dependent. The total efficiency
\begin{equation}
    \eta \equiv \eta_e \eta_n \equiv \frac{\varDelta\EM}{\varDelta\ES}\frac{\varDelta N_\text{actions}}{N_\text{re}}
\end{equation}
quantifies the fraction of energy input into the system via charged Source structures that is converted into charged Machine structures.\\[4mm]
The Supplemental Material contains details on the model, discussions of the dynamics of the reaction network and the navigation of parameter space using differentiable programming, and a comparison of timescales for Machine activation and spontaneous dissociation. It further includes detailed analyses of the presented Langevin Dynamics simulations, a discussion of the behavior of additional designs, and an interpretation of the performance for the examples of a synthetic pump and a synthetic walker, which includes Refs.~\cite{bolh25, gill00, kram40, hang90, tryg04, cerj81, wale04, maur05, shes95, bitz06, jax2018github, scho21, blon22, bayd18, rume86, weng64}. Our work is based on the software packages JAX and JAX-MD~\cite{jax2018github,scho21}, with built-in hardware acceleration and ensemble vectorization.

\begin{acknowledgments} 
    We thank Edouard Hannezo, Ella King, Maximilian Lechner, and J\'er\'emie Palacci for stimulating discussions, and Edouard Hannezo, Maximilian Hübl, and Maitane Muñoz-Basagoiti for helpful comments on the manuscript. This research was funded in part by the Austrian Science Fund (FWF) [\href{https://doi.org/10.55776/PAT8537123}{10.55776/PAT8537123}].
\end{acknowledgments}

\bibliography{literature}

\appendix

\section{Model details}

\begin{figure}[tb]
    \includegraphics[width=0.49\textwidth]{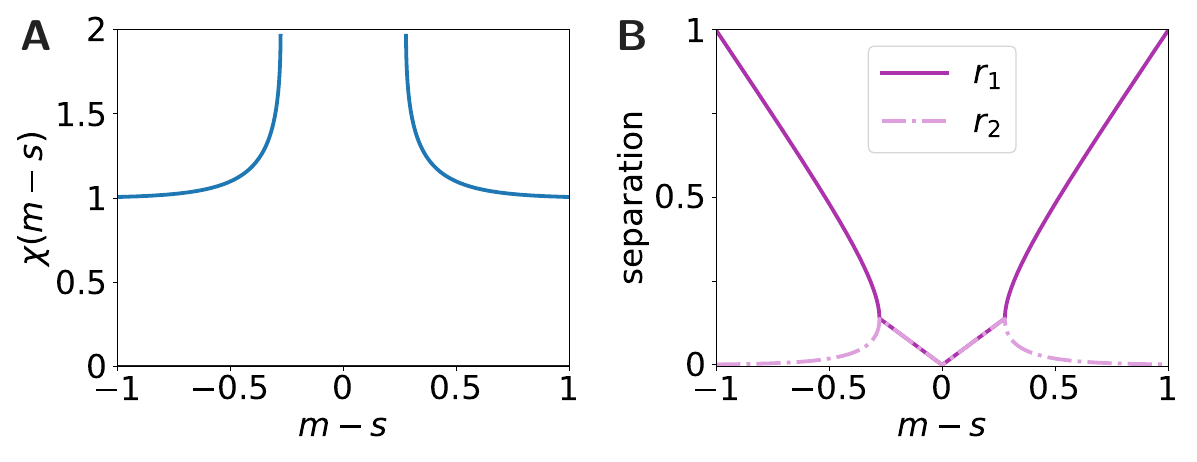}\centering
    \caption{Calculating the separations $r_1$ and $r_2$ of collinear dimers for an interaction range of $\alpha=5$ used in this work. \textsf{(A)} The function $\chi(m-s)$ from Eq.~\eqref{eq:chi} is strictly real and positive when $|m-s| > \ln4/\alpha$. \textsf{(B)} The separations from Eqs.~\eqref{eq:r1} and \eqref{eq:r2} are plotted as function of the length difference $m-s$ of Machine and Source. They are identical when $|m-s| \leq \ln4/\alpha$.}
    \label{fig:cases}
\end{figure}

The full energy of the system as function of dimer lengths $m,s$ and separations $r_1,r_2$ is given by 
\begin{equation}\label{eq:Etot_si}
    E(m, s, r_1, r_2) = E_\text{M}(m) + E_\text{S}(s) + E_\mathrm{int}(r_1) + E_\mathrm{int}(r_2).
\end{equation}
Here, the energy profiles of the Machine and Source, $E_\text{M}(m)$ and $E_\text{S}(s)$, respectively, are given by the function
\begin{align}\label{eq:Er}
    E(r) &= \epsilon_\uu \left(e^{-2\alpha_\uu (r - r_\uu)} - 2e^{-\alpha_\uu (r - r_\uu)}\right) + cr \nonumber\\ 
         &\quad + \epsilon_\cc \left(e^{-2\alpha_\cc (r_\cc - r)} - 2e^{-\alpha_\cc (r_\cc - r)}\right),
\end{align}
with $r$ replaced by either $m$ or $s$ and where the parameters $\epsilon_\uu$, $\alpha_\uu$, $r_\uu$, $\epsilon_\cc$, $\alpha_\cc$, $r_\cc$, and $c$ are different for the two energy profiles. 
The attractive interactions between Machine and Source ends are given by
\begin{equation}\label{eq:Vint}
    E_\text{int}(r) = \epsilon\left(e^{-2\alpha r} - 2e^{-\alpha r}\right).
\end{equation}
The actual values of these 16 parameters $\theta$ are listed in Table~I in the main text. For fixed $m$ and $s$, we can analytically calculate the minimum energy configuration
\begin{equation}
    E_\text{min}(m,s) \equiv \min_{r_1,\,r_2} E(m,s,r_1,r_2).
\end{equation}
This minimum always occurs when the dimers are collinear, with the ends of the shorter dimer located at or within the longer dimer. This means that
\begin{equation}
    |m - s| = r_1 + r_2
\end{equation}
and we can therefore set $r_1=|m-s|-r_2$ to find stable configurations as function of the separation $r_2$ by solving
\begin{equation}
    \nabla_{r_2} E(m,s,r_2) = \nabla_{r_2} \big[E_\text{int}(|m-s|-r_2) + E_\text{int}(r_2)\big] = 0.
\end{equation}
The resulting, general expressions of the separations are
\begin{equation}\label{eq:r1}
    r_1 = |m-s| - r_2
\end{equation}
and
\begin{align}\label{eq:r2}
    r_2 = 
    \begin{cases}
        |m-s|\,/\,2, & \text{if } e^{\alpha|m-s|} \leq 4, \\ 
        \ln\chi/\alpha,  & \text{if } e^{\alpha|m-s|} > 4,
    \end{cases}
\end{align}
where
\begin{equation}\label{eq:chi}
    \chi \equiv \frac{1}{2}\left(e^{\alpha|m-s|} - e^{\alpha|m-s|/2} \sqrt{e^{\alpha|m-s|}-4}\right).
\end{equation}
Note that $\chi$ is strictly real and positive when $|m-s| > \ln4/\alpha$.
For an interaction range of $\alpha=5$ used in this work, the expressions for $\chi$ and the separations are plotted in Fig.~\ref{fig:cases} as function of the length difference $m-s$ of Machine and Source.

\section{Dynamics of the reaction network} 

\begin{figure}[tb]
    \includegraphics[width=0.49\textwidth]{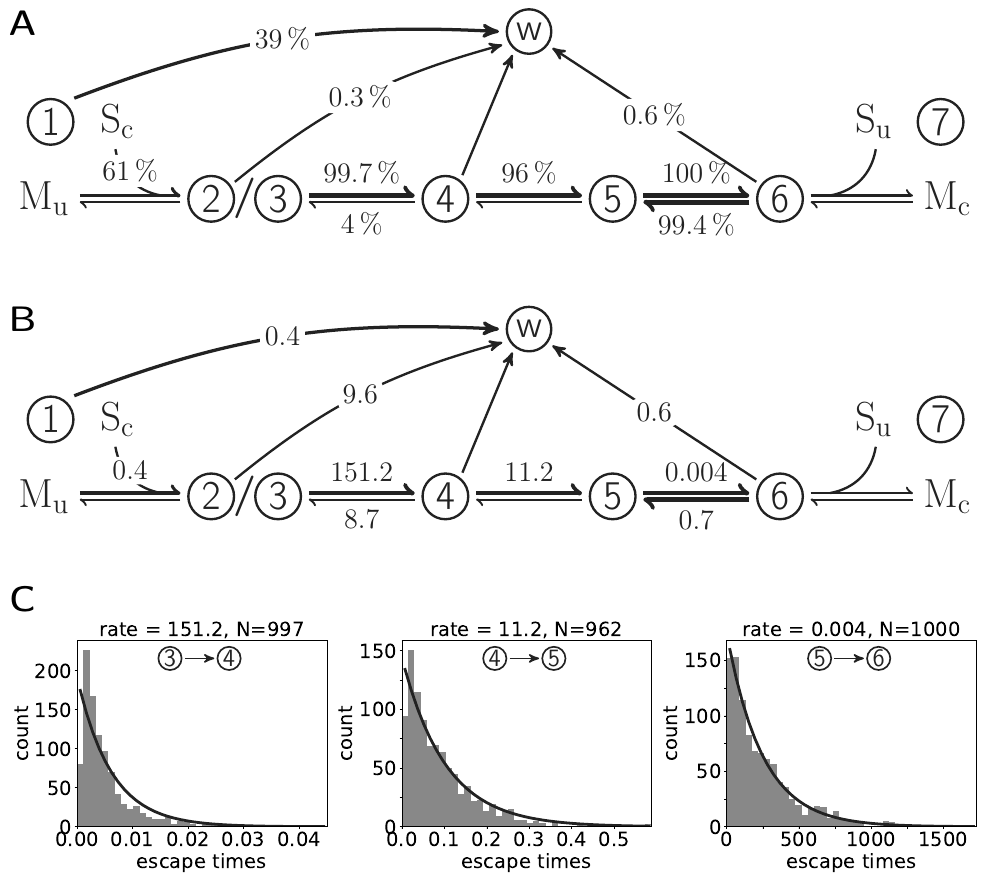}\centering
    \caption{Transition network between meta-stable states in our model with transition rates measured in ensemble BD simulations, where the state labels refer to Fig.~2\textsf{I} and Fig.~4 of the main text. We measure transition rates by running the dynamics of an ensemble of $1000$ simulations, initialized in states \textcircled{\textsf{\scriptsize 1}} to \textcircled{\textsf{\scriptsize 6}}, and record into which state they transition at the measured first-passage time, for the design with parameters $\theta_1(\epsilon=25)$. For each observed transition, we state the corresponding fraction of observations (\textsf{A}) and the transition rate (\textsf{B}) inferred from an exponential fit to the distribution of first-passage times (\textsf{C}) or the mean value for small fractions. The thickness of the arrows illustrates the flow of the network according to these fractions. \textsf{(C) The activating transition \textcircled{\textsf{\scriptsize 3}} $\to$ \textcircled{\textsf{\scriptsize 4}} is roughly ten times faster than the energy-delivery transition \textcircled{\textsf{\scriptsize 4}} $\to$ \textcircled{\textsf{\scriptsize 5}}, while the dissociation transition \textcircled{\textsf{\scriptsize 5}} $\to$ \textcircled{\textsf{\scriptsize 6}} is significantly slower than all other transitions.} Transitions to the waste state are irreversible and undesirable. This network does not consider the recharging of the Source or the spontaneous discharging of the free Machine.}
    \label{fig:network_rates}
\end{figure}

The transition network between meta-stable states in our model is illustrated in Fig.~\ref{fig:network_rates}.
The state labels refer to Fig.~2\textsf{I} and Fig.~4 of the main text and the uncharged and charged states of Machine and Source are denoted with $\text{M}_\text{u/c}$ and $\text{S}_\text{u/c}$. 
By treating the system as a continuous time Markov chain with absorbing states \textcircled{\textsf{\scriptsize 7}} and \textcircled{\textsf{\scriptsize w}}, we can solve this problem analytically if we either calculate or measure all reaction rates. However, proper individual rate calculations are notoriously complicated, especially at high binding energies, which is the regime in which our mechanism is most effective. 

\subsection{Measuring transition rates in BD simulations}
In order to measure the transition rates, we run ensemble Brownian Dynamics (BD) simulations (see details in Section~\ref{sec:bd}) with a batch size of $1000$, initialized in states \textcircled{\textsf{\scriptsize 1}} to \textcircled{\textsf{\scriptsize 6}}, and record the state they transition into at the measured first-passage time. Figure~\ref{fig:network_rates} presents the transition network stating the fraction of observations for each observed transition and the transition rate inferred from an exponential fit to the distribution of first-passage times or the mean value for small fractions for the design with parameters $\theta_1(\epsilon=25)$. The thickness of the arrows illustrates the flow of the network according to the observed fractions.
Identification of all meta-stable states in the transition network is not trivial and is highly dependent on the thresholds used to determine bound states and dimer lengths. We use different thresholds for different initial states: if dimer ends are initially bound we use a threshold of $E_\text{int} < 0.05\epsilon$ to identify an unbound state, and a threshold of $E_\text{int} > 0.95\epsilon$ if they are initially unbound. We take the length of the opposite state as length thresholds, \ie, $m > m_\cc$ and $s > s_\uu$ for initial states \textcircled{\textsf{\scriptsize 1}} to \textcircled{\textsf{\scriptsize 4}}, and $m < m_\uu$ and $s < s_\cc$ for initial states \textcircled{\textsf{\scriptsize 5}} and \textcircled{\textsf{\scriptsize 6}}. We present data for the transition network at high binding energy, where transitions are the clearest to the best of our knowledge. Note that transitions can go back and forth between neighboring states.

The observed activating transition \textcircled{\textsf{\scriptsize 3}} $\to$ \textcircled{\textsf{\scriptsize 4}} is roughly ten times faster than the energy-delivery transition \textcircled{\textsf{\scriptsize 4}} $\to$ \textcircled{\textsf{\scriptsize 5}}, while the dissociation transition \textcircled{\textsf{\scriptsize 5}} $\to$ \textcircled{\textsf{\scriptsize 6}} is significantly slower than all other transitions. We do not observe complete dissociation of Machine and Source at the high binding energy of $\epsilon=25$ for a maximum simulation time of $10^9$. Note that we did not include the transition \textcircled{\textsf{\scriptsize 6}} $\to$ \textcircled{\textsf{\scriptsize w}} in Fig.~5 of the main text even though we observe this transition rarely, as we consider state \textcircled{\textsf{\scriptsize 6}} as a successful event already in the definition of the performance. We do not discriminate waste states with one or zero bonds between Source and Machine here. If one cares about the structures being fully dissociated, we investigate the behavior of a stricter definition of the performance in Fig.~\ref{fig:si_performance}.
However, this is a simplistic method to estimate rates, which is not at all exact. Proper estimation of transition rates is an active field of research. Using techniques as those developed by Bolhuis~\cite{bolh25} and others, one could potentially calculate the rates more accurately, but it is beyond the scope of this paper.

\subsection{Calculating transition rates using Kramers transition state theory}

The dynamics of a rate-based reaction network consisting of $N$ species with concentrations $x_i$ can be described by a deterministic reaction-rate equation using the law of mass action in the thermodynamic limit and with a linear noise approximation~\cite{gill00}
\begin{equation}
    \frac{\partial x_i}{\partial t} = \sum_{j=1}^N \big( k_{ji}x_j - k_{ij}x_i \big).
\end{equation}
Transition state theory developed by Kramers allows to calculate reaction rates $k_{ij}$ between two meta-stable states that are connected by an energy barrier $\varDelta E_{it}$ along a reaction coordinate~\cite{kram40, hang90}
\begin{equation}\label{eq:kramers}
    k_{ij} = \frac{\omega_\uu^t}{2\pi\gamma}\frac{\prod_l \omega_l^i}{\prod_{l'} \omega_{l'}^t} e^{-\varDelta E_{it} / \kT}.
\end{equation}
The prefactor includes the friction coefficient $\gamma$ and information about the curvatures of the underlying energy landscape. The frequency $\omega_\uu^t = \sqrt{-\lambda_\uu^t}$ corresponds to the unstable vibrational mode at the transition state given by the negative eigenvalue of the Hessian of the energy, and $\{\omega_{l'}^t\}$ and $\{\omega_l^i\}$ are the frequencies of the stable modes at the transition state and the metastable state, respectively. 

\begin{table}[t]
    \centering
    \begin{tabular}{cccccc}
        \toprule
        design & transition & $\varDelta E$ & Kramers rate & BD rate & error \\
        \midrule
        $\theta_1(\epsilon=25)$ & \textcircled{\textsf{\scriptsize 3}} $\to$ \textcircled{\textsf{\scriptsize 4}} & $0.24$ & $305.2$ & $151.2$ & $102\,\%$ \\
        Fig.~\ref{fig:network_rates} & \textcircled{\textsf{\scriptsize 4}} $\to$ \textcircled{\textsf{\scriptsize 5}} & $3.30$ & $8.3$ & $11.2$ & $26\,\%$ \\
        \midrule
        $\theta_1(\epsilon=20)$ & \textcircled{\textsf{\scriptsize 3}} $\to$ \textcircled{\textsf{\scriptsize 4}} & $1.21$ & $297.8$ & $48.6$ & $513\,\%$ \\
        & \textcircled{\textsf{\scriptsize 4}} $\to$ \textcircled{\textsf{\scriptsize 5}} & $3.24$ & $9.5$ & $13.1$ & $27\,\%$ \\
        \midrule
        $\theta_1(\epsilon=15)$ & \textcircled{\textsf{\scriptsize 3}} $\to$ \textcircled{\textsf{\scriptsize 4}} & $2.66$ & $90.8$ & $16.6$ & $447\,\%$ \\
        & \textcircled{\textsf{\scriptsize 4}} $\to$ \textcircled{\textsf{\scriptsize 5}} & $3.12$ & $12.4$ & $82.4$ & $85\,\%$ \\
        \bottomrule
    \end{tabular}
    \caption{Comparison of theoretical predictions of transition rates to measurements from ensemble BD simulations for the activating and energy-delivery reactions in the design $\theta_1$ for three different binding energies. The calculated rates applying Kramers transition state theory differ significantly from the measured rates as shown by the relative error. Rates with lower energy barriers $\varDelta E$ differ more as Kramers theory becomes less accurate.}
    \label{tab:rates}
\end{table}

The Hessian of the energy has five zero modes for the local minima and transition states of the collinear transitions \textcircled{\textsf{\scriptsize 3}} $\to$ \textcircled{\textsf{\scriptsize 4}} $\to$ \textcircled{\textsf{\scriptsize 5}}, corresponding to the three translational and two rotational degrees of freedom of the collinear dimers. Applying Kramers theory yields the transition rates stated in Table~\ref{tab:rates}. Even for the energy-delivery rate $k_{45}$, which is arguably the clearest transition in the network, the relative error is $26\,\%$. 
The activating transition \textcircled{\textsf{\scriptsize 3}} $\to$ \textcircled{\textsf{\scriptsize 4}} has a smaller energy barrier and the error in rates is much higher, in agreement with Kramers theory becoming less accurate for small energy barriers. For the second design with different energy barriers the errors behave as expected.

We attempted to optimize the behavior of the system using Kramers transition state theory to estimate all relevant rates and simplistic assumptions about association and dissociation rates, as discussed in the next section. However, there are several complications, which ended up making this approach not feasible. The discrepancy between theoretical predictions and rates measured in ensemble BD simulations demonstrates that we cannot use a Markov model to optimize the system. Kramers transition state theory breaks down if the energy barriers are small, which is the case for essential transitions in desired parameter regimes in which the mechanism works as intended. Additionally, the landscape changes during optimization with states appearing and disappearing when changing parameters. Even though our model is flexible using a general method to calculate mean first-passage times, it caused problems.

\begin{figure}[tbp]
    \includegraphics[width=0.49\textwidth]{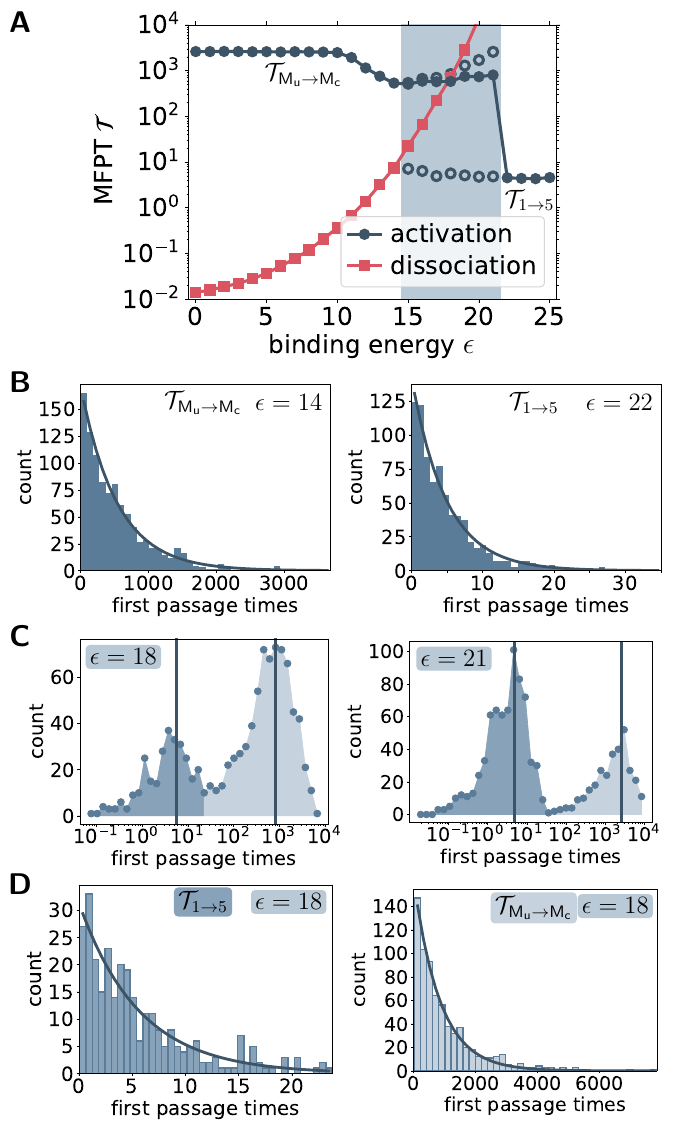}\centering
    \caption{Comparison of mean first-passage times (MFPTs) for activation of the Machine and spontaneous dissociation of a Machine-Source complex as function of the binding energy. \textsf{(A)} At low binding energies $\epsilon < 15$, only the slow pathway $\mathcal{T}_{\Mu\to \Mc}$ of the Machine transitioning on its own is relevant, and the dynamics is limited by the Machine activation energy. At high binding energies $\epsilon > 21$, only the fast pathway $\mathcal{T}_{1\to5}$ following the coupled transition is relevant and the dynamics is diffusion limited. In these two regimes with only one relevant activation pathway, MFPTs are extracted from exponential fits to the distribution of first passage times (\textsf{B}). In the intermediate regime with $15\leq\epsilon\leq21$ (gray region in \textsf{A}), both activation pathways are relevant as shown by the histograms on logarithmically spaced bins to capture both timescales \textsf{(C)}. We extract MFPTs from two separate exponential fits for both pathways individually (\textsf{D}), shown as open circles in \textsf{A}, and approximate $\mathcal{T}_\text{activation} \approx w\mathcal{T}_{1\to5} + (1-w)\mathcal{T}_{\Mu\to \Mc}$ as weighted mean. The spontaneous dissociation curve increases super-exponentially as there are two dissociation events. The MFPT for activation $\mathcal{T}_{1\to 5}$ matches the spontaneous dissociation time at intermediate binding energies $\epsilon \approx 14$. The value for high binding energies $\mathcal{T}_\text{activation}(\epsilon=25) \simeq 4.6$ suggests that an active dissociation mechanism operating on that timescale would be most effective.}
    \label{fig:mfpts_comp}
\end{figure}

\subsection{Comparison of timescales for Machine activation and spontaneous dissociation}
In order to compare the times needed for association and energy-delivery transitions to take place with the time for spontaneous dissociation, we perform ensemble BD simulations to estimate the mean first-passage times (MFPTs) for activation of the Machine in the presence of the Source and for the spontaneous dissociation of a Machine-Source complex. They are presented in Fig.~\ref{fig:mfpts_comp}\textsf{A} for the design with parameters $\theta_1$ as functions of the binding energy. For measuring $\mathcal{T}_\text{activation}$ at a given $\epsilon$, we prepare an ensemble of 1000 BD simulations in state \textcircled{\textsf{\scriptsize 1}} and evolve each one until the Machine is activated. Note that we do not consider transitions to the waste state in this analysis. For measuring $\mathcal{T}_\text{dissociation} \equiv \mathcal{T}_{5\to7}$, we prepare the ensemble in the doubly-bound Machine-Source complex \textcircled{\textsf{\scriptsize 5}} and evolve it until the dissociated state \textcircled{\textsf{\scriptsize 7}} is reached. We extract $\mathcal{T}_\text{dissociation}(\epsilon)$ from the exponential distribution of first passage times by fitting an exponentially decaying function.

The measured spontaneous dissociation curve increases super-exponentially because there are two dissociation events, one for each bond. If it were only one dissociation event, this would instead scale exponentially with the binding energy.
The behavior of the activation curve $\mathcal{T}_\text{activation}(\epsilon)$ is more nuanced. At low binding energies, $\epsilon < 15$, only the slow pathway $\mathcal{T}_{\Mu\to \Mc}$ of the Machine transitioning on its own is relevant, and the dynamics is limited by the Machine activation energy. The baseline value of $\mathcal{T}_\text{activation}(\epsilon=0) \simeq 2610$ agrees roughly with the theoretical prediction of $1 / k_{\uu\cc} \simeq 2440$ from Fig.~\ref{fig:machine}. At high binding energies $\epsilon > 21$, only the fast pathway $\mathcal{T}_{1\to5}$ following the coupled transition is relevant and the dynamics is diffusion limited: energy delivery happens fast, but it takes time for the dimers to associate (\cf rates in Fig.~\ref{fig:network_rates}\textsf{B}). In both regimes, the histograms of first passage times follow exponential distributions as presented in Fig.~\ref{fig:mfpts_comp}\textsf{B} and we extract $\mathcal{T}_\text{activation}$ from an exponential fit.

Interestingly, we observe that $\mathcal{T}_\text{activation}$ is roughly constant for $\epsilon\leq10$ and drops significantly for $10\leq\epsilon\leq15$. This is consistent with the observed increase of the measures of merit shown in Fig.~6 of the main text.
At intermediate binding energies $15\leq\epsilon\leq21$ (gray region in Fig.~\ref{fig:mfpts_comp}\textsf{A}), both pathways are relevant as presented in Fig.~\ref{fig:mfpts_comp}\textsf{C} by histograms on logarithmically spaced bins to capture both timescales for two example binding energies. Therefore, we extract MFPTs from two separate exponential fits for both pathways individually as shown in Fig.~\ref{fig:mfpts_comp}\textsf{D} for $\epsilon=18$, which correspond to the open circles in \textsf{A}. Interestingly, the timescale $\mathcal{T}_{1\to5}$ decreases very slowly if the transition occurs for large $\epsilon$, but the likelihood of this pathway increases with increasing $\epsilon$. We approximate $\mathcal{T}_\text{activation} \approx w\mathcal{T}_{1\to5} + (1-w)\mathcal{T}_{\Mu\to \Mc}$ as weighted mean using the relative likelihood $w$, which is shown as full circles in Fig.~\ref{fig:mfpts_comp}\textsf{A}.
At the largest binding energy we have $\mathcal{T}_\text{activation}(\epsilon=25)\simeq4.6$, suggesting that an active dissociation mechanism operating on a roughly comparable timescale would be most effective, as it allows the energy-delivery mechanism to take place before causing dissociation and still be significantly faster than spontaneous dissociation. This agrees roughly with the peak of the performance around $\tau_\dd\approx1$ in Fig.~6\textsf{D} of the main text. Furthermore, we can observe where $\mathcal{T}_{1\to 5}$ matches the spontaneous dissociation time as the binding energy decreases. This occurs roughly around binding energies of $\epsilon \approx 14$, again consistent with the observed peak in the performance in Fig.~6\textsf{A} of the main text for $\tau_\dd=\infty$ and giving insight into the importance of dissociation in our mechanism.

\section{Navigating parameter space with differentiable programming}

\subsection{General optimization framework}

We have developed a flexible and adaptable differentiable state-based model to navigate the complex design space. 
We proceed with the following five steps:

\emph{(Step 1)} Identify a set of relevant states, defined as local minima in the energy landscape for a given set of parameters. This can be done by hand or through an automated process. 

\emph{(Step 2)} Calculate transition pathways using the doubly-nudged elastic band (DNEB) method~\cite{tryg04}, which is an efficient double-ended method to estimate transition states. We then refine this estimate using eigenvector following~\cite{cerj81, wale04, maur05}.

\emph{(Step 3)} Use transition state theory~\cite{kram40,hang90} to approximate rates of all forward and backward transitions, setting the Kramers prefactor from Eq.~\eqref{eq:kramers} to 1 for simplicity.

\emph{(Step 4)} Treating the system as a linear Markov chain, with transition probabilities $p_{ij} = k_{ij}\varDelta t$ given by the transition rates times a small time step, calculate the mean first-passage time by solving a system of linear equations~\cite{shes95} 
\begin{align}
    \mathcal{T}_{i\to j} = 1 + \sum_{k\neq j} p_{ik}\mathcal{T}_{k\to j}.
\end{align}

\emph{(Step 5)} Define a loss function that describes a particular behavior we want to optimize. For example, we can define the loss function as the ratio of mean first-passage times to the usable charged state of the coupled system compared to equilibrium without the Source,
\begin{equation}\label{eq:loss}
    \mathcal{L}(\theta) = \frac{\mathcal{T}_\text{usable}^\text{non-eq}(\theta)}{\mathcal{T}_{\Mu\to\Mc}^\text{eq}(\theta)} + \mathcal{C}(\theta),
\end{equation}
where $\theta$ is the set of all relevant parameters and $\mathcal{C}(\theta)$ are physical constraints on \eg energy barriers, lengths, and curvatures.
By implementing steps 2-5 using Automatic Differentiation~\cite{bayd18, rume86, weng64}, we can calculate the gradient $\nabla_\theta \mathcal{L}$, and feed this gradient into standard gradient-based optimization routines. 
Specifically, we minimize $\mathcal{L}$ using the standard L-BFGS algorithm with a learning rate of $l=10^{-3}$. 
We recalculate the network topology every ten steps, but in every step we refine our estimates of the positions of the minima (using FIRE energy minimization~\cite{bitz06}) and transition states (using eigenvector following~\cite{wale04}). This model is implemented using the software packages JAX and JAX-MD~\cite{jax2018github, scho21}, with built-in hardware acceleration and ensemble vectorization.

\subsection{Concrete example illustrating the optimization framework}

\begin{figure}[tb]
    \includegraphics[width=0.49\textwidth]{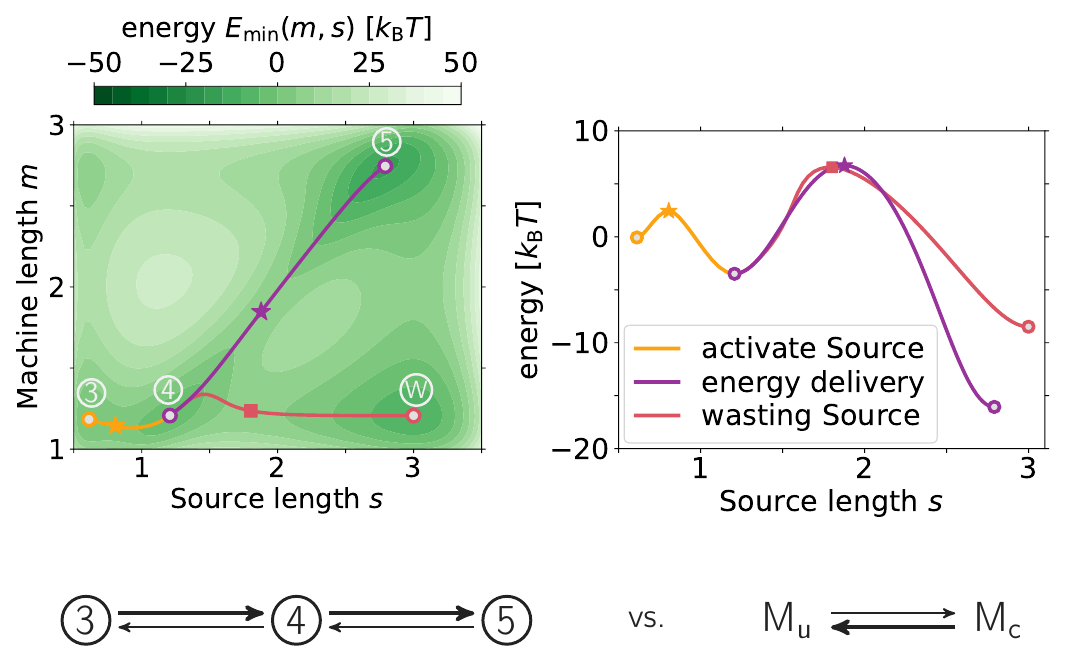}\centering
    \caption{Illustration of our differentiable state-based model. We identify all relevant local minima in the energy landscape and calculate forward and backward transition rates to optimize the dynamics of the Machine along the desired coupled transition pathway \textcircled{\textsf{\scriptsize 3}} $\to$ \textcircled{\textsf{\scriptsize 4}} $\to$ \textcircled{\textsf{\scriptsize 5}} using the Source, compared to a Machine on its own, $\Mu\to\Mc$. The thickness of the arrows illustrates preferred directions based on energy barriers.}
    \label{fig:si_opt_states}
\end{figure}

As an example, we illustrate how we can use this model to adjust the energy profiles starting from those in Fig.~2\textsf{D} and arriving at those in Fig.~2\textsf{F} in the main text. 
Starting with the energy landscape shown in Fig.~\ref{fig:si_opt_states} and in Fig.~2\textsf{D-E} in the main text, we identify the local minima along the coupled transition pathway to be states \textcircled{\textsf{\scriptsize 3}}, \textcircled{\textsf{\scriptsize 4}}, and \textcircled{\textsf{\scriptsize 5}}.
This can be compared to a Machine on its own, without a Source, which can (slowly) transition back and forth between the charged and uncharged states. 
We calculate all transition rates for the two networks
\begin{align}
    \large{\textcircled{\normalsize \textsf{3}}} \xrightleftharpoons[k_{43}]{k_{34}} \large{\textcircled{\normalsize \textsf{4}}} \xrightleftharpoons[k_{54}]{k_{45}} \large{\textcircled{\normalsize \textsf{5}}}, && \Mu \xrightleftharpoons[k_\text{cu}]{k_\text{uc}} \Mc,
\end{align}
and get the mean first-passage times through the relations
\begin{align}\label{eq:mfpt23}
    \mathcal{T}_{3\to 5} = \frac{1 + \dfrac{p_{34}}{p_{43}+p_{45}}}{p_{34} - \dfrac{p_{34}p_{43}}{p_{43}+p_{45}}}, && \mathcal{T}_{\Mu\to \Mc}^\text{eq} = \frac{1}{2p_{\Mu\to \Mc}}.
\end{align}
We define the loss 
\begin{equation}\label{eq:loss_opt_si}
    \mathcal{L}(\theta) = \frac{\mathcal{T}_{3\to5}(\theta)}{\mathcal{T}_{\Mu\to \Mc}^\text{eq}(\theta)} + \mathcal{C}(\theta),
\end{equation}
with physical constraints
\begin{align}\label{eq:constraints}
    \mathcal{C}(\theta) &= c(E_\text{M}^\text{a} - 10) + c(E_\text{S}^\text{a} - 10) \nonumber\\
    &\quad + c(E_\text{M}^\text{a} - \varDelta E_\text{M} - 8) + c(E_\text{S}^\text{a} + \varDelta E_\text{S} - 8) \nonumber\\
    &\quad + c(\varDelta E_\text{M} - 5) + c(30 - \varDelta E_\text{S}) + c(s_\cc - 0.5) \nonumber\\
    &\quad + c(2 - (s_\uu - m_\cc)) + c(s_\uu - m_\cc - 0.3)
\end{align}
on energy and length differences (\emph{cf.} Fig.~1 in the main text), using a one-sided spring function as penalty
\begin{equation}
    c(x) = \begin{cases}
               x^2, & \text{if } x<0,\\
               0, & \text{if } x\geq0.
    \end{cases}
\end{equation}
The purpose of the constraints is to keep the optimizer in physical regimes with desired behavior. Forward energy barriers are at least $10\,\kT$, backward energy barriers at least $8\,\kT$, the required Machine energy capacity is $5\,\kT$ and the Source energy capacity is not larger than $30\,\kT$ to prevent inefficient solutions. Lengths are in physical regimes with a minimal length for the charged Source as well as bounds for the difference of uncharged Source and charged Machine lengths.

\begin{figure}[tb]
    \includegraphics[width=0.49\textwidth]{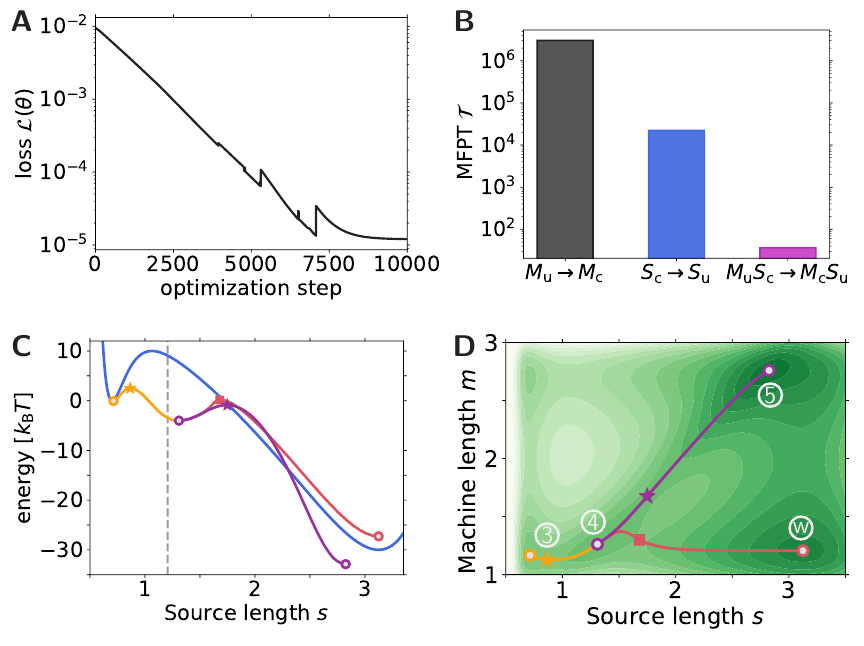}\centering
    \caption{Optimization results of the collinear dimer model with physical constraints on energy barriers and lengths. (\textsf{A}) Loss from Eq.~\eqref{eq:loss_opt_si} during optimization with constraints from Eq.~\eqref{eq:constraints} using the L-BFGS optimizer from JAXopt~\citep{blon22}. The optimization parameters $\theta$ are given by the Source energy profile and the binding energy. Starting from an initial value $\mathcal{L}(\theta_\text{ini})\simeq9.5\cdot10^{-3}$ the loss decreases by almost three orders of magnitude until it converges at an optimized value $\mathcal{L}(\theta_\text{opt})\simeq1.2\cdot10^{-5}$ after $10^4$ optimization steps with a final magnitude of the loss gradient of $|\nabla_\theta\mathcal{L}(\theta_\text{opt})|\simeq1.2\cdot10^{-2}$. (\textsf{B}) Mean first-passage times of Machine and Source alone as well as for the coupled reaction of the optimized system. The mean first-passage time of the coupled reaction is significantly smaller compared to the equilibrium systems. (\textsf{C}) Energy profiles as functions of the Source length for optimized parameters with energy barriers of $2.6\,\kT$ for the activating transition (orange curve), $3.1\,\kT$ for the energy delivery transition (purple curve), and $4.2\,\kT$ for the wasted transition (red curve). The corresponding energy pathways are calculated using the DNEB method. (\textsf{D}) Energy landscape with transition pathways. The coupled collinear states $\Mu\Sc$ and $\Mc\Su$ correspond to states \textcircled{\textsf{\scriptsize 3}} and \textcircled{\textsf{\scriptsize 5}}. The corresponding optimized parameters $\theta_1$ are listed in Table~I of the main text.}
    \label{fig:SI_optimization}
\end{figure}

For simplicity, we keep parameters corresponding to the Machine energy profile and the interaction range $1/\alpha$ fixed and optimize with respect to parameters of the Source energy profile and the binding energy $\epsilon$. The optimization results are shown in Fig.~\ref{fig:SI_optimization}.
Starting from an initial value $\mathcal{L}(\theta_\text{ini})\simeq9.5\cdot10^{-3}$ the loss decreases by almost three orders of magnitude until it converges at an optimized value $\mathcal{L}(\theta_\text{opt})\simeq1.2\cdot10^{-5}$ after $10^4$ optimization steps with a final magnitude of the loss gradient of $|\nabla_\theta\mathcal{L}(\theta_\text{opt})|\simeq1.2\cdot10^{-2}$. The discrete jumps in the loss function occur due to recalculation of the energy landscape every 10 optimization steps by recalculating all DNEB pathways from a set of initial guesses for the minima. The mean first-passage time of the coupled reaction is significantly smaller compared to the equilibrium systems. 
Prominent modifications in the Source energy profile are the increased energy capacity and the shift of the Source transition state to lower $s\simeq1.0$. Both together have the desired effect of making the Source energy profile steeper near $s\approx1.2$. Interestingly, the optimization did not increase the binding energy much, with a final value of $\epsilon\simeq15.3\,\kT$.
This enables us to significantly reduce the purple energy barrier to $3.1\,\kT$ as shown in Fig.~\ref{fig:SI_optimization}\textsf{C} and Fig.~2\textsf{F} in the main text. We find that both the orange barrier for activating the Source and the purple barrier for energy delivery are of roughly the same height after the optimization. Note that the transition to the waste state \textcircled{\textsf{\scriptsize w}} (red curve) does not enter into the loss function in this example, though more sophisticated versions of this can do so.

The main limitations of this approach are the approximations and assumptions inherent in transition state theory. Our goal is to have small energy barriers for the energy delivery pathway, but this is exactly where transition state theory becomes less accurate as demonstrated above. In addition, we frequently encounter topology changes during the optimization when states appear and disappear (\eg from Fig.~2\textsf{C} to Fig.~2\textsf{E} in the main text), which causes problems for training. This approach also completely fails, for example, in cases like the one shown in Fig.~6\textsf{I} in the main text. 
Therefore, we do not claim to have found properly optimized designs, and instead use it as a guide to navigate this complex parameter space.

\section{Overdamped Langevin Dynamics simulations}\label{sec:bd}

The dynamics of two dimers with coordinates $\bm{r}_i$ in the energy landscape $V(\bm{r}_i,\bm{r}_j)$ from Eq.~\eqref{eq:Etot_si} is governed by an overdamped Langevin equation
\begin{equation}\label{eq:Langevin}
    \gamma \frac{\dd\bm{r}_i}{\dt} = -\sum_{j\neq i}\bm{\nabla}_i V(\bm{r}_i, \bm{r}_j) + \sqrt{2\gamma\kT}\bm{\xi}_i(t),
\end{equation}
with friction coefficient $\gamma=0.1$, thermal energy $\kT=1$, and white noise $\bm{\xi}_i(t)$. We simulate the dimers in a three-dimensional box of side length 10 with periodic boundary conditions, starting from a random configuration initialized with the dimers separated and in their closed state. We run Brownian Dynamics (BD) using a time step $\dt=10^{-5}$ for $2\cdot10^{10}$ steps such that we observe multiple Machine actions. For each data point in Figs.~6 and 7 in the main text, we run either 200 (Fig.~6\textsf{A-C}) or ten (Fig.~6\textsf{D-F}) such simulations with different seeds and analyze them over time windows $T_\text{sim}$ from the first to the last Machine charging event. This removes bias caused by the initial conditions.
In order to replicate the effect of a constant concentration of usable charged Source structures, we introduce a recharging scheme that converts an uncharged Source structure into a charged Source structure. Recharging only occurs when the Source and Machine are sufficiently separated, which is when both dimer ends are outside a sphere of radius 
\begin{equation}
 r_\text{re} = \frac{s_\uu-s_\cc}{2} + \frac{8}{\alpha}.
\end{equation}
Thus, we drive the system out of equilibrium resulting in a non-equilibrium steady state.
Note that while this directly injects energy into the system, the coupled energy-delivery reaction itself is a purely dissipative process that is driven only by the initial conditions. For simulations with finite $\tau_\dd$, recharging the Source resets the binding strength $\epsilon$ to its initial $\epsilon_0$.

\subsection{Machine on its own}

\begin{figure}[tb]
    \includegraphics[width=0.49\textwidth]{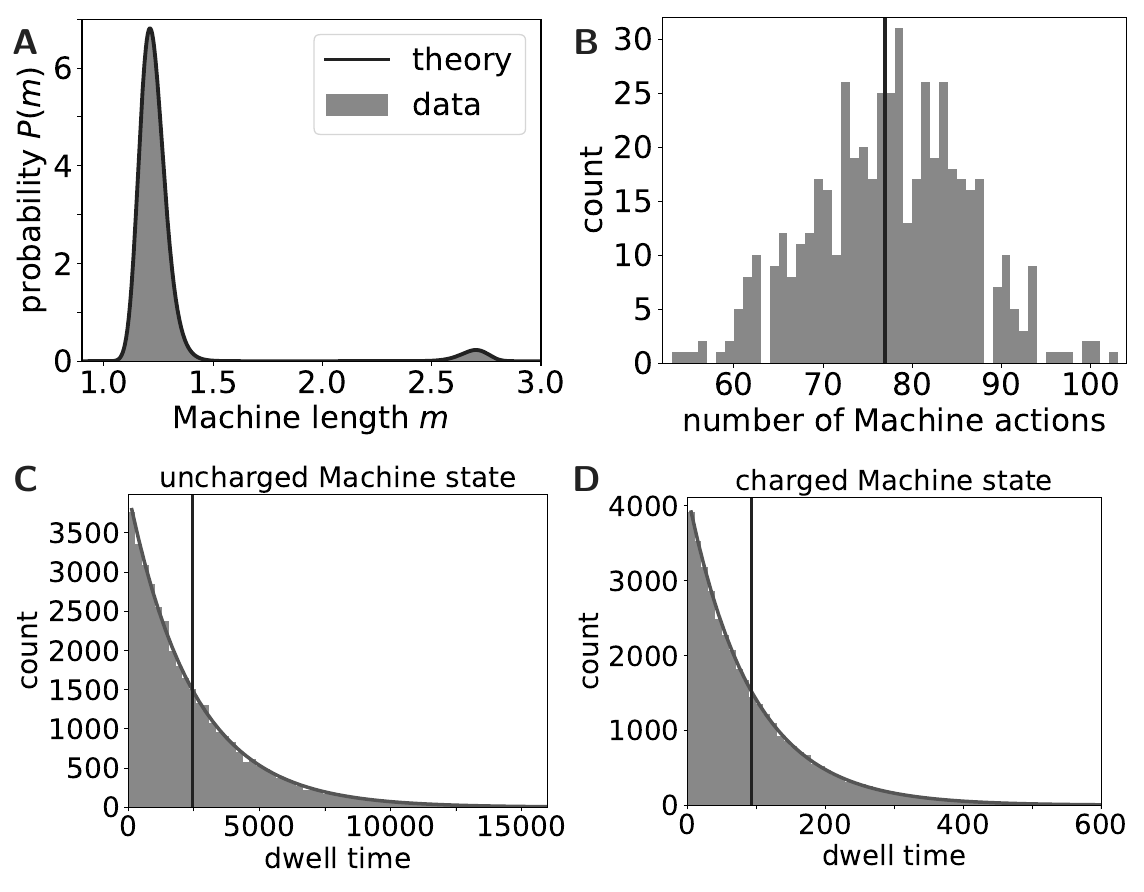}
    \caption{Analysis of the Machine equilibrium simulation.
    \textsf{(A)} Histogram of Machine lengths saved every $10^4$ simulation steps. The data reveals a perfect agreement to the Boltzmann equilibrium probability distribution including entropic contributions, defined in Eqs.~\eqref{eq:boltzmann} and \eqref{eq:free_energy}. \textsf{(B)} Histogram of completed Machine actions during the simulation, in which the Machine transitions from its uncharged state to its charged state and back. The mean value is given by the black line at $N_\text{actions}^\text{Machine only}\simeq76.94 \pm 8.55$ with standard deviation. \textsf{(C)} Histogram of dwell times in the uncharged Machine state. The black line shows the mean value, corresponding to a transition rate of $k_{\uu\cc}\simeq(4.10\pm0.18)\cdot10^{-4}$ and the gray curve is an exponential fit with inferred rate $k_{\uu\cc}^\text{fit}\simeq4.03\cdot10^{-4}$. \textsf{(D)} Histogram of dwell times in the charged Machine state. The inferred transition rates are $k_{\cc\uu}\simeq(1.08\pm0.05)\cdot10^{-2}$ and $k_{\cc\uu}^\text{fit}\simeq1.08\cdot10^{-2}$. We used 100 long simulation runs with different seeds in \textsf{A} and 500 in \textsf{B-D}. We use the Machine parameters $\theta_1$ listed in Table~I of the main text.}
    \label{fig:machine}
\end{figure}

We present an analysis of the Machine equilibrium simulation in Fig.~\ref{fig:machine}. 
The normalized histogram of Machine lengths reveals a perfect agreement to the Boltzmann equilibrium probability distribution
\begin{equation}\label{eq:boltzmann}
    P_\text{eq}(m) = \frac{1}{Z} e^{-\beta F(m)},
\end{equation}
with the normalization constant $Z=\int P_\text{eq}(m)\dd m$. The free energy $F(m)$ of a particle escape in three dimensions that includes entropic contributions is given by
\begin{equation}\label{eq:free_energy}
  F(r) = E(r) - TS(r) = E(r) - 2\kT\ln(r),
\end{equation}
where the energy $E(r)$ is given by Eq.~\eqref{eq:Er}. The performed simulations validate the expressions in Eqs.~\eqref{eq:boltzmann} and \eqref{eq:free_energy} numerically. Across 500 long simulation runs (details in Methods) we observe $N_\text{actions}^\text{Machine only}\simeq76.94 \pm 8.55$ complete Machine actions, with the Machine transitioning from its uncharged state to the charged state and back, as shown by the histogram in Fig.~\ref{fig:machine}\textsf{B}. The histograms of dwell times in the uncharged and charged Machine states show exponentially decaying times in agreement with Kramers theory. The extracted values of the transition rates between both states are $k_{\uu\cc}\simeq(4.10\pm0.18)\cdot10^{-4}$ and $k_{\cc\uu}\simeq(1.08\pm0.05)\cdot10^{-2}$. They are in good agreement with the corresponding rates obtained from Kramers theory, $4.28\cdot10^{-4}$ and $1.10\cdot10^{-2}$, applying Eq.~\eqref{eq:kramers} using the free energy from Eq.~\eqref{eq:free_energy}.

\subsection{Machine and Source}

\subsubsection{Machine actions and Source recharging events}
\begin{figure}[htb]
    \includegraphics[width=0.49\textwidth]{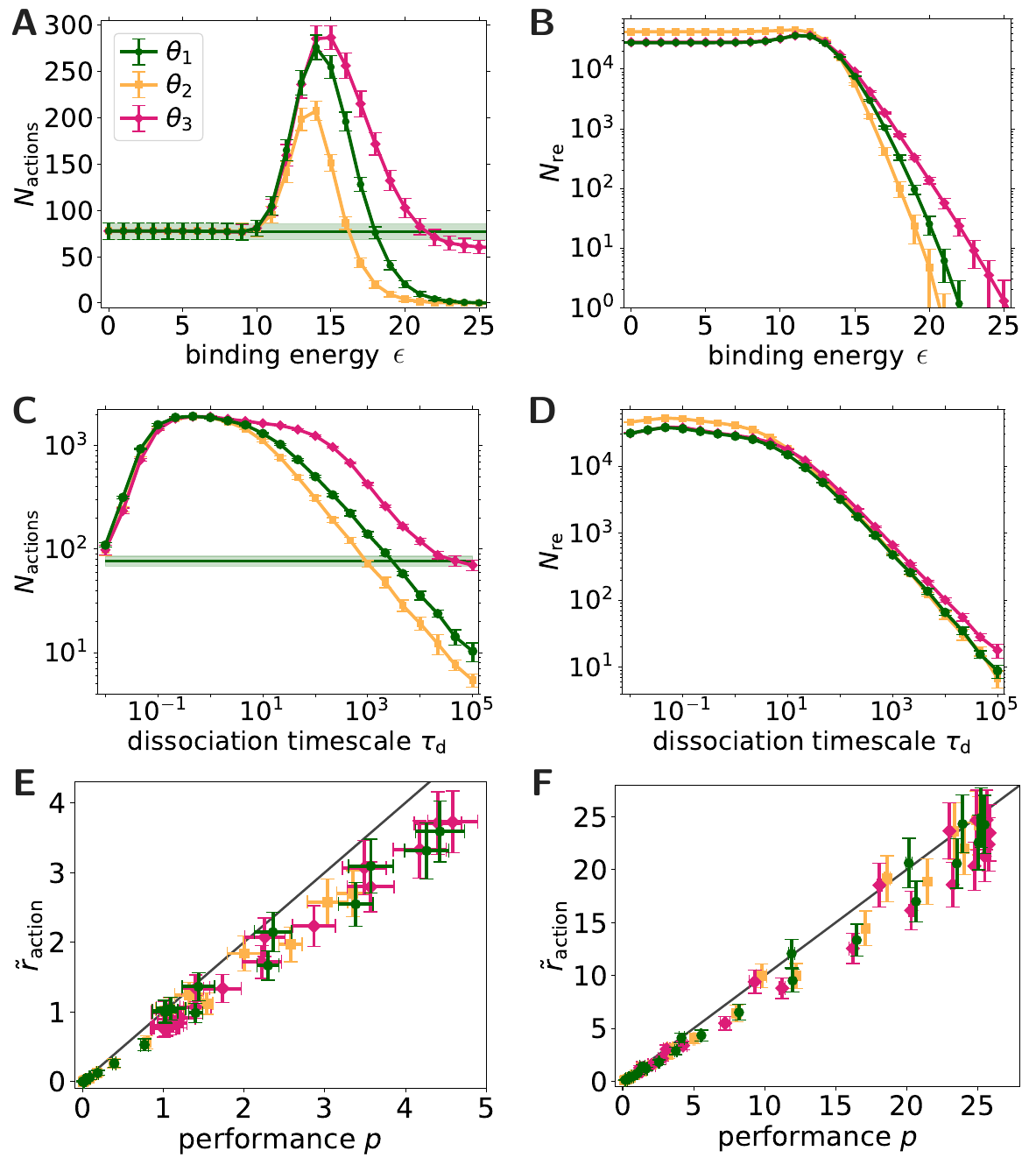}
    \caption{Number of Machine actions $N_\text{actions}$ and number of Source recharging events $N_\text{re}$ in the designs reported in Fig.~6 of the main text as functions of the binding energy $\epsilon$ and the dissociation timescale $\tau_\dd$. The number of Machine actions peaks at a given $\epsilon$ or $\tau_\dd$ and drops for higher binding energies or dissociation timescales similar to the measures of merit. The horizontal lines in \textsf{A} and \textsf{C} indicate the equilibrium values $N_\text{actions}^\text{Machine only}$ of the Machine without the Source (mean value with standard deviation visualized by the shaded region). (\textsf{E} and \textsf{F}) Comparison of the performance and the relative action rate $\tilde{r}_\mathrm{action}$ from Eq.~\eqref{eq:si_action_rate}, which is calculated for the data presented in \textsf{A} and \textsf{C}.}
    \label{fig:Nactions_Nre}
\end{figure}

We now turn to the long BD simulations of the coupled system in the designs reported in Fig.~6 in the main text. The number of Machine actions $N_\text{actions}$ and the number of Source recharging events $N_\text{re}$ are shown in Fig.~\ref{fig:Nactions_Nre}.
The number of Machine actions peaks at a given $\epsilon$ or $\tau_\dd$ and drops for higher binding energies or dissociation timescales similar to the measures of merit. It drops to 0 for the first two designs as the system is stuck in the complex, while it drops to the equilibrium value for the third design, in which the system is left with only one bond after the energy-delivery reaction and the Machine transitions between its two states. The horizontal lines indicate the equilibrium values $N_\text{actions}^\text{Machine only}$ of the Machine without the Source and the difference to the nonequilibrium values of the coupled system gives the number of additional Machine charging events 
\begin{equation}
    \varDelta N_\text{actions} \equiv N_\text{actions}^\text{with Source} - N_\text{actions}^\text{Machine only}.
\end{equation}
The number of Source recharging events plateaus for small binding energies and dissociation timescales at values $N_\text{re}>10^4$ and shows an exponential decay for high $\epsilon$ and a power-law decay for high $\tau_\dd$.

We calculate the relative action rate 
\begin{equation}\label{eq:si_action_rate}
    \tilde{r}_\mathrm{action} \equiv N_\text{actions}^\text{with Source} \;/\; N_\text{actions}^\text{Machine only}
\end{equation}
for the data presented in Fig.~\ref{fig:Nactions_Nre}\textsf{A} and \textsf{C} and compare it to the performance for varying $\epsilon$ and $\tau_\dd$ in \textsf{E} and \textsf{F}. The data shows that the performance gives a good estimate for the relative action rate.

\subsubsection{Progression of the energy-delivery mechanism}

\begin{figure}[htb]
    \includegraphics[width=0.45\textwidth]{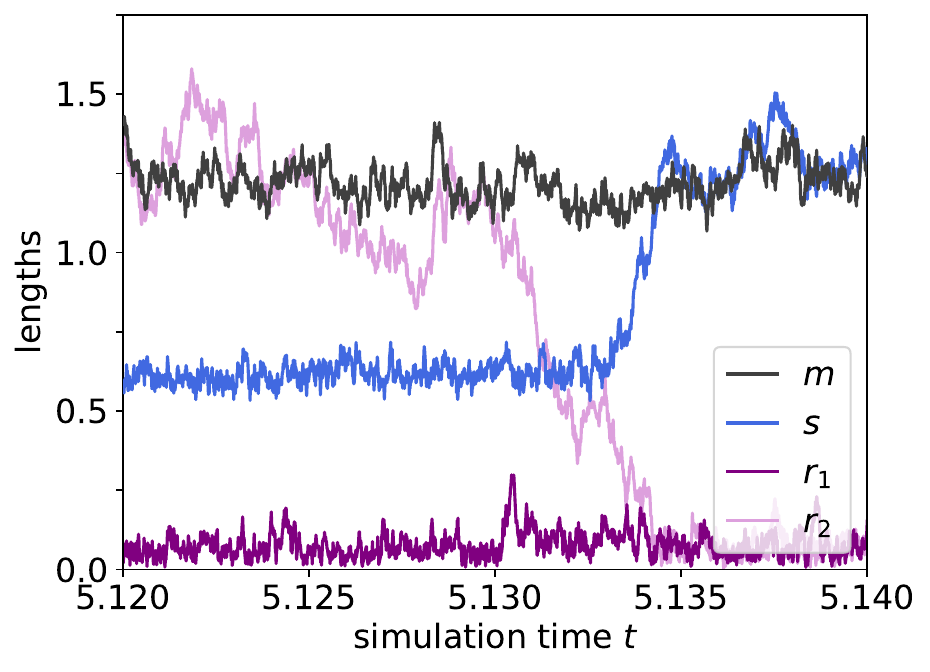}\centering
    \caption{Detailed version of the very fast transition from state \textcircled{\textsf{\scriptsize 2}} to state \textcircled{\textsf{\scriptsize 4}} shown in Fig.~4 of the main text. The lifetime of the intermediate state \textcircled{\textsf{\scriptsize 3}}, in which the Machine activates the Source, is so small that we sometimes have difficulty resolving it.}
    \label{fig:SI_MD_detail}
\end{figure}

A more detailed version of the fast transition from state \textcircled{\textsf{\scriptsize 2}} to \textcircled{\textsf{\scriptsize 4}} as reported in Fig.~4 of the main text is shown in Fig.~\ref{fig:SI_MD_detail}.
The lifetime of the intermediate state \textcircled{\textsf{\scriptsize 3}}, in which the Machine activates the Source, is so small that we have difficulty resolving it.

\begin{figure*}[htbp]
    \includegraphics[width=\textwidth]{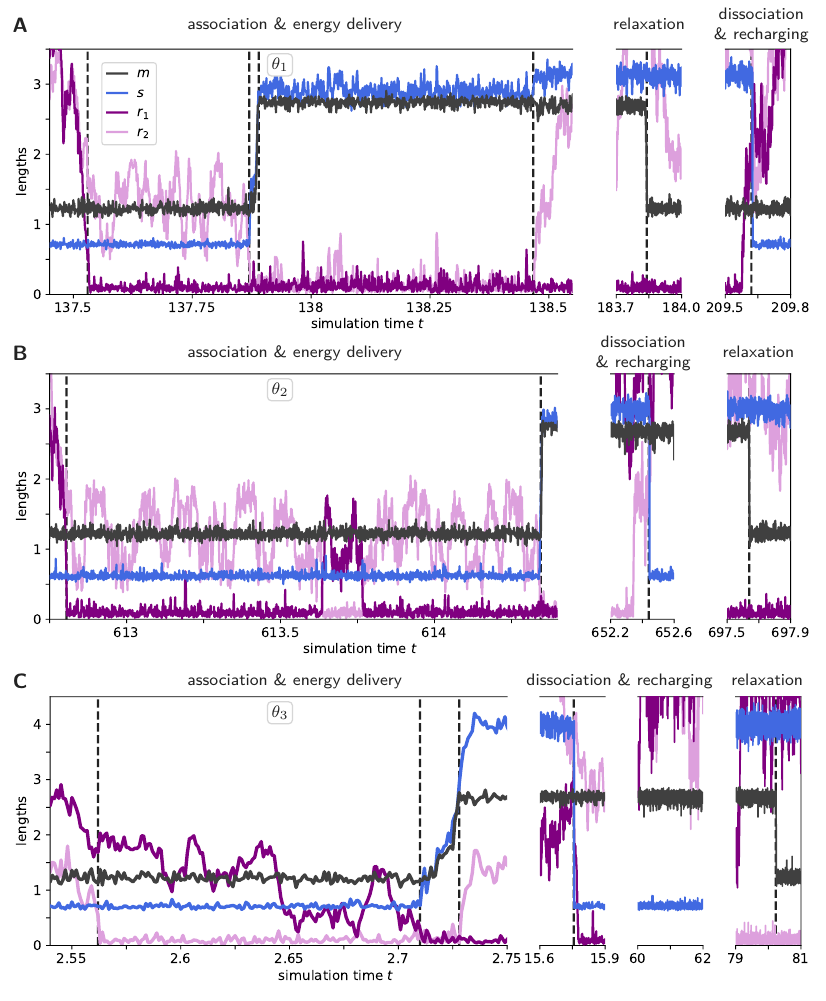}\centering
    \caption{Example trajectories from Langevin Dynamics simulations without forced dissociation for three designs $\theta_1$, $\theta_2$, and $\theta_3$ with binding energy $\epsilon=15$. The time evolution of all four relevant lengths, Machine and Source lengths, $m$ and $s$, as well as the dimer-end separations $r_1$ and $r_2$, is shown for a complete Machine action, demonstrating the progression of the energy-delivery mechanism. The examples demonstrate a fast coupled energy-delivery transition once Machine and Source have associated. While the Machine in \textsf{A} is singly-bound to the Source when spontaneously relaxing to its ground state followed by Source recharging, the Machines in \textsf{B} and \textsf{C} spend time in the usable charged state completely dissociated from the Source before relaxation (see the second-to-last panels in \textsf{B} and \textsf{C}). The dashed vertical lines mark important time points in which the state of the coupled system changes. Note that some reversible transitions in intermediate periods are omitted in this figure to make it comprehensible.}
    \label{fig:SI_MD_examples}
\end{figure*}

Example trajectories from Langevin Dynamics simulations without forced dissociation for three designs $\theta_1$, $\theta_2$, and $\theta_3$ with binding energy $\epsilon=15$ are presented in Fig.~\ref{fig:SI_MD_examples}, in comparison to Fig.~4 in the main text. The example trajectory in Fig.~\ref{fig:SI_MD_examples}\textsf{A} for $\theta_1$ starts with forming a single bond between Machine and Source. Once both bonds are formed, the coupled energy delivery transition happens extremely quickly. The system stays in the doubly-bound complex until one bond breaks. The Machine spontaneously transitions to its ground state while bound to the Source at one side. The Source is recharged once both bonds are broken and the structures are sufficiently separated.
The energy-delivery mechanism for $\theta_2$ shown in Fig.~\ref{fig:SI_MD_examples}\textsf{B} happens even faster. In this example trajectory, the Machine stays in its usable charged state dissociated from the Source for a while before it relaxes to its ground state.
The trajectory for design $\theta_3$ shown in Fig.~\ref{fig:SI_MD_examples}\textsf{C} demonstrates a quick progression to the singly-bound complex after the coupled energy-delivery transition, which is a consequence of the increased difference in open lengths of Machine and Source. We again observe a time period of a completely unbound Machine in its usable charged state before relaxing to its ground state.

\subsubsection{Steady-state probability distributions}
\begin{figure*}[htbp]
    \includegraphics[width=\textwidth]{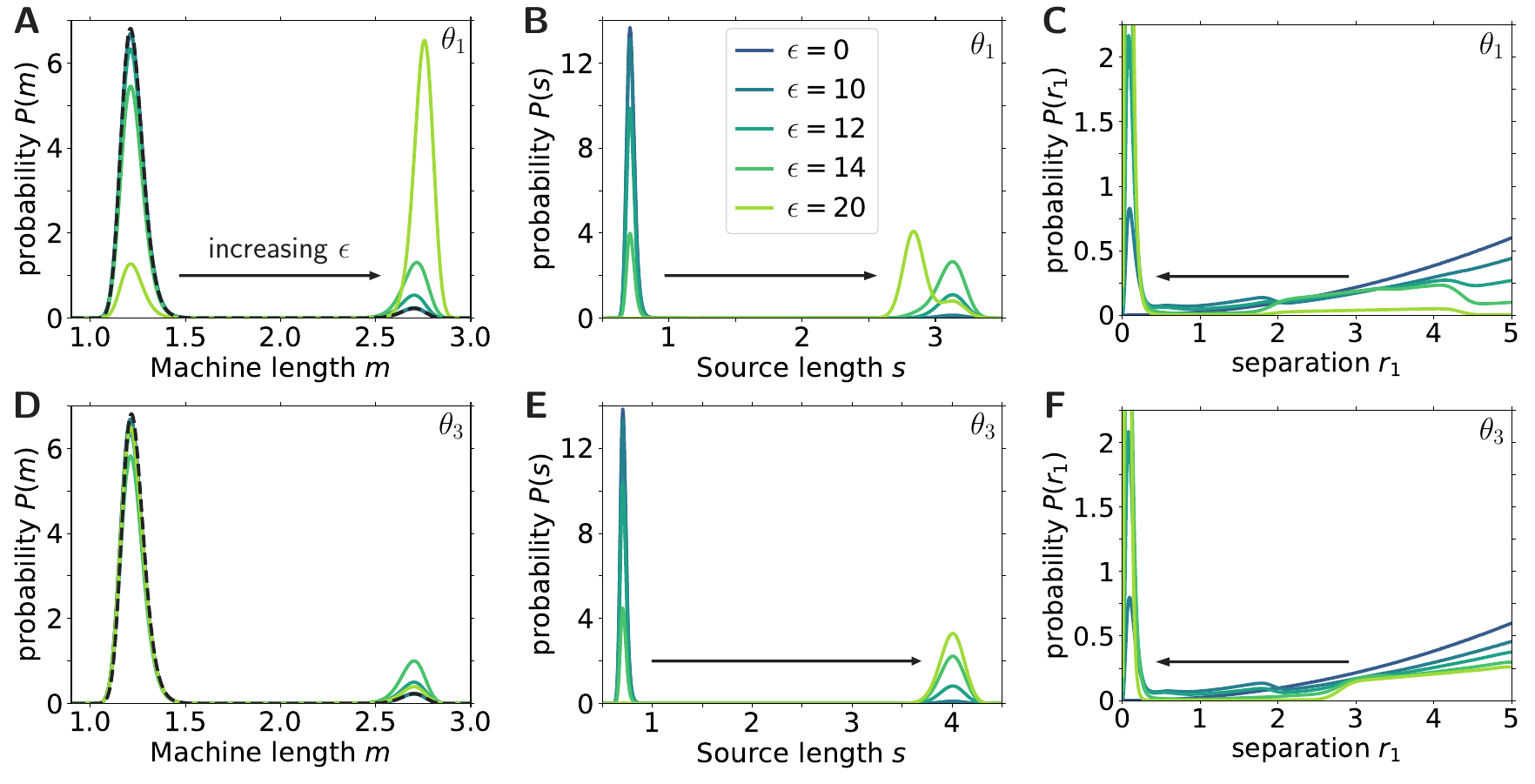}\centering
    \caption{Steady-state probability distributions $P(m)$, $P(s)$, and $P(r_1)$ for increasing binding energies $\epsilon=[0,10,12,14,20]$ extracted from long simulations of the coupled system for the green design in Fig.~6 of the main text with parameters $\theta_1$. \textsf{(A)} The data demonstrates a strong occupation enhancement of the target, high-energy Machine state at $m\approx2.7$ for increasing $\epsilon$. It matches the theoretical prediction given by the Boltzmann equilibrium probability distribution $P_\text{eq}$ including entropic contributions as defined in Eqs.~\eqref{eq:boltzmann} and \eqref{eq:free_energy} (dashed line) for small $\epsilon$. \textsf{(B)} We observe an occupation enhancement of the uncharged Source state at $s\approx3.0$. The distribution for low $\epsilon$ reveals the nonequilibrium driving of the system with many Source recharging events (see Fig.~\ref{fig:Nactions_Nre}), as otherwise the Source would spent most of the time in the uncharged, low-energy state. For $\epsilon=20$ the Source is predominantly found in the complex state with $s\simeq m_\cc < s_\uu$. \textsf{(C)} The distribution of separations is shifted to shorter distances for increasing binding energy, demonstrating that the system is driven to the complex state of bound Machine and Source. We analyze data until half the simulation box length $L/2=5$. We only show the probability of one separation ($r_1$), but we verified that indeed both distributions are identical, $P(r_1)=P(r_2)$, due to the symmetry of the model. Altogether, the data shows a tradeoff: the occupation enhancement of the usable Machine state comes at the cost of forming a complex with the Source. 
    \textsf{(D-F)} Same analysis as in \textsf{A-C}, but extracted from simulations of the magenta design in Fig.~6 of the main text with parameters $\theta_3$. The occupation enhancement of the target Machine state is less prominent; it increases up to $\epsilon \simeq 14$, but decreases again for $\epsilon = 20$. Since the charged Machine state is much smaller than the uncharged Source state for this data, the complex is only bound at one end (shown in \textsf{F}), allowing the Machine to transition back to the uncharged state.
    All distributions are calculated from histograms of the trajectory and are normalized such that their integral gives one.}
    \label{fig:probabilities}
\end{figure*}

Figure~\ref{fig:probabilities}\textsf{A-C} shows the steady-state probability distributions $P(m)$, $P(s)$, and $P(r_1)$ for increasing binding energies $\epsilon=[0,10,12,14,20]$ extracted from long simulations of the coupled system for the green design in Fig.~6 of the main text with parameters $\theta_1$. The Machine length distribution matches the Boltzmann equilibrium probability distribution $P_\text{eq}$ (dashed line) from Eqs.~\eqref{eq:boltzmann} and \eqref{eq:free_energy} for small $\epsilon$ and demonstrates a strong occupation enhancement of the target, high-energy Machine state at $m\approx2.7$ for increasing $\epsilon$. The distribution $P(s)$ for low $\epsilon$ reveals the nonequilibrium driving of the system with many Source recharging events (see Fig.~\ref{fig:Nactions_Nre}), as otherwise the Source would spent most of the time in the uncharged, low-energy state. We also observe an occupation enhancement of the uncharged Source state at $s\approx3.0$ for high $\epsilon$. For $\epsilon=20$ the Source is predominantly found in the complex state with $s\simeq m_\cc < s_\uu$. The distribution of separations $P(r_1)$ is shifted to shorter distances for increasing binding energy, demonstrating that the system is driven to the complex state of bound Machine and Source. Altogether, the data shows a tradeoff: the occupation enhancement of the usable Machine state comes at the cost of forming a complex with the Source. Figure~\ref{fig:probabilities}\textsf{D-F} shows the steady-state probability distributions extracted from simulations of the magenta design in Fig.~6 of the main text with parameters $\theta_3$. The occupation enhancement of the target Machine state is less prominent; it increases up to $\epsilon \simeq 14$, but decreases again for $\epsilon = 20$. Since the charged Machine state is much smaller than the uncharged Source state for this data, the complex is only bound at one end (shown in \textsf{F}), allowing the Machine to transition back to the uncharged state.

\subsubsection{Strict performance requiring full dissociation}
\begin{figure}[htb]
    \includegraphics[width=0.49\textwidth]{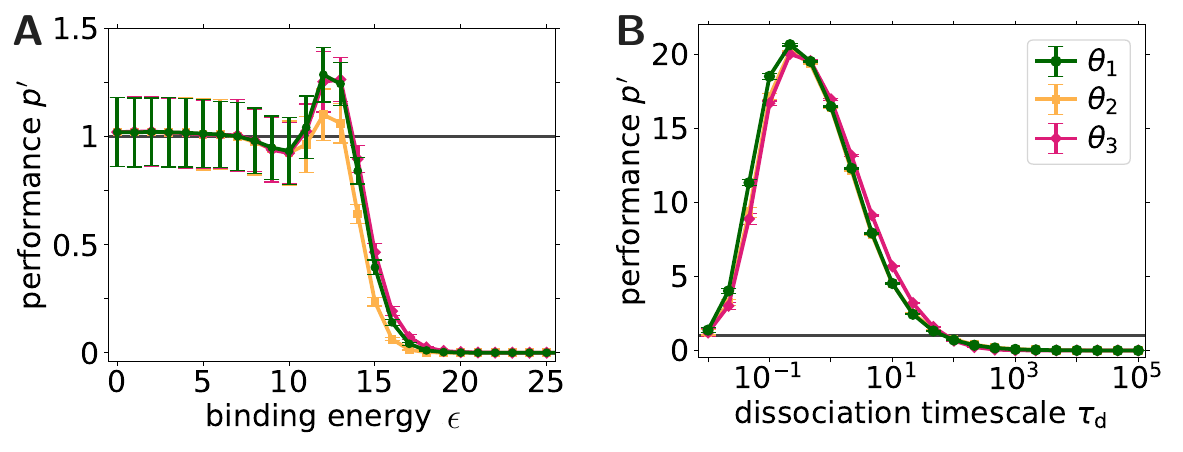}
    \caption{Performance of the designs reported in Fig.~6 of the main text as function of the binding energy $\epsilon$ and the dissociation timescale $\tau_\dd$ for the strict definition of a usable charged state that requires the Source to be completely detached from the Machine. (\textsf{A}) The peak around $\epsilon\simeq12$ with $p' \simeq 1.3 \pm 0.1$ for designs $\theta_1$ and $\theta_3$ is statistically significant above 1 (indicated by the horizontal line), demonstrating a slight performance increase even for the strict definition. (\textsf{B}) The non-monotonic behavior is similar to the moderate performance discussed in the main text, but the peak is shifted to smaller $\tau_\dd$, slightly decreased with $p'\approx16.42$ at $\tau_\dd=1$, and it drops faster with increasing $\tau_\dd$.}
    \label{fig:si_performance}
\end{figure}

The performance $p \equiv P_\text{usable}^\text{with Source} \,/\, P^\text{Machine only}_\text{usable}$ measures the increase in likelihood that, at any given time, the Machine is in its usable charged state, compared to equilibrium. When $p>1$, the Machine spends more time in the usable state in the presence of the Source than in equilibrium and the mechanism is deemed successful. Conversely, $p<1$ means that the Source is actually harmful, for example by forming long-lived Machine-Source complexes. 
We also consider a stricter definition of the performance that requires the Source to be completely detached. Its behavior as a function of $\epsilon$ and $\tau_\dd$ is shown in Fig.~\ref{fig:si_performance}.
The mechanism works over roughly four decades in $\tau_\mathrm{d}$ with $p'\approx16.42$ for the simulation shown in Fig.~4 of the main text, and is statistically significant above 1 with $p' \simeq 1.3 \pm 0.1$ for the green design at constant binding energy $\epsilon=12\,\kT$. This demonstrates a slight performance increase even for the strict definition of the performance. However, the strict performance drops faster than the moderate version in which we include singly-bound states and it is below one for $\epsilon=15\,\kT$.

We can distinguish input and output power of the mechanism. The input power $\mathcal{P}_\text{in}$ is given by the number of recharging events during the simulation times the energy capacity of the Source. The output power $\mathcal{P}_\text{out}$ measures the excess number of Machine transitions during the simulation multiplied by the energy capacity of the Machine,
\begin{align}
    \mathcal{P}_\text{in} \equiv \frac{N_\text{re}}{T_\text{sim}}\varDelta E_\text{S}, &&
    \mathcal{P}_\text{out} \equiv \frac{\varDelta N_\text{actions}}{T_\text{sim}}\varDelta E_\text{M}.
\end{align}
The output power measures work that can be done by the Machine on the timescale of the usable state. The input power measures the energy injected to the system on that timescale. Together with the efficiency they obey the relation $\eta = \mathcal{P}_\text{out} / \mathcal{P}_\text{in}$.

\begin{figure*}[htbp]
    \includegraphics[width=\textwidth]{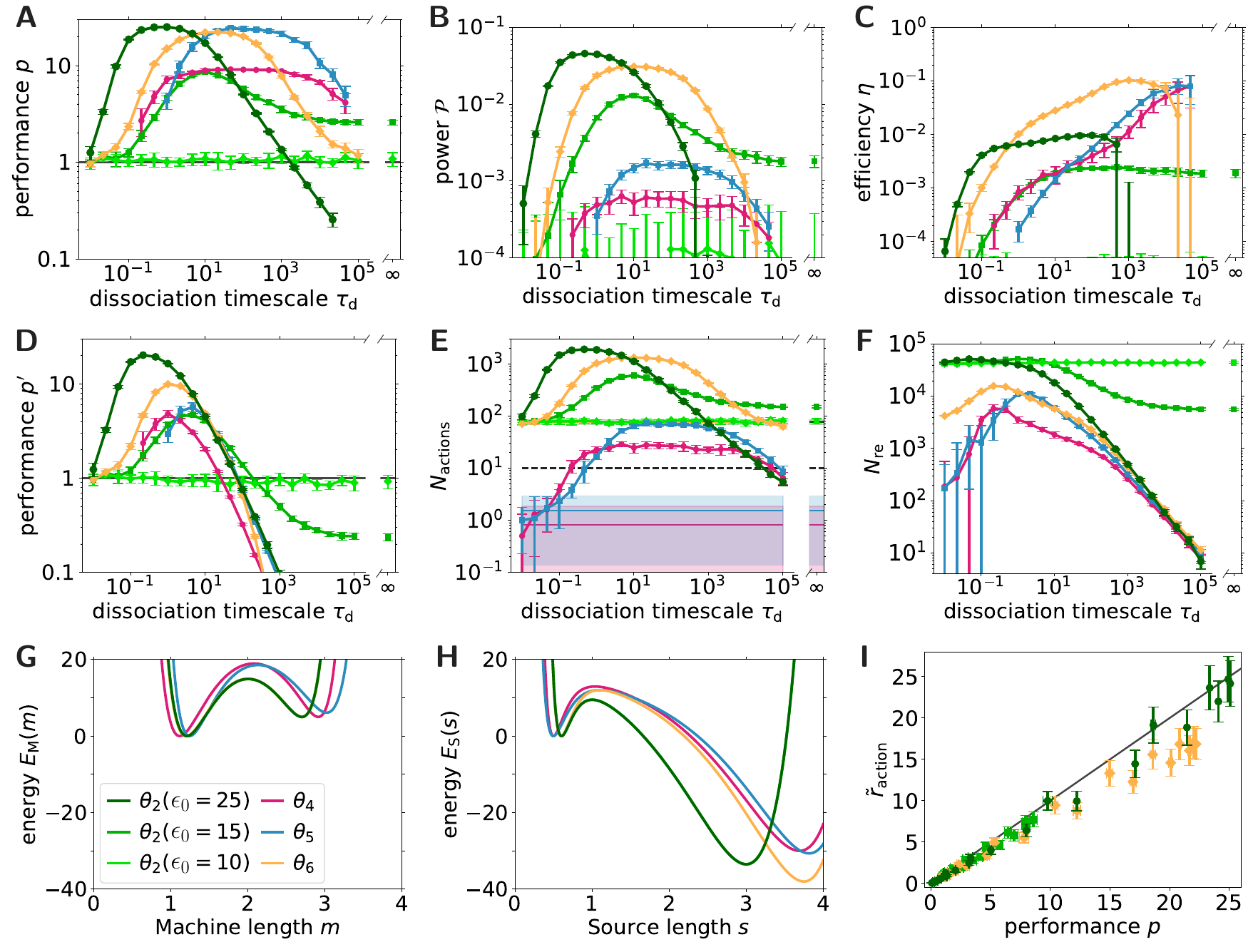}\centering
    \caption{Measures of merit for different designs as function of the dissociation timescale $\tau_\dd$ in addition to those analyzed in the main text. The data demonstrates robustness of the energy-delivery mechanism over a broad range of dissociation timescales and across various designs as well as nontrivial tradeoffs between these measures when changing the energy profiles or $\tau_\dd$. The solid horizontal lines in \textsf{E} indicate the equilibrium values $N_\text{actions}^\text{Machine only}$ of the Machine without the Source (mean value with standard deviation visualized by the shaded region). The dashed line at $N_\text{actions}=10$ shows the threshold for reporting data in \textsf{A-D}. The green designs differ only in the maximum binding energy $\epsilon_0$ and correspond to the orange design presented in Fig.~6 of the main text. The corresponding model parameters are listed in Table~\ref{tab:params_si}. \textsf{(I)} Comparison of the performance and the relative action rate $\tilde{r}_\text{action}$ from Eq.~\eqref{eq:si_action_rate}. The performance is slightly larger than the relative action rate leading to the data deviating slightly from the straight line indicating equality. Designs $\theta_4$ and $\theta_5$ are not shown as they only have a few Machine actions for the given simulation time due to their large Machine activation barriers (\cf \textsf{E} and \textsf{G}).}
    \label{fig:si_measures}
\end{figure*}

\begin{table*}[htb]\centering
  \setlength{\tabcolsep}{3pt}
  \begin{tabular}{lllllllllllllllll}
   \toprule
    & \multicolumn{7}{l}{Machine} & \multicolumn{7}{l}{Source} & \multicolumn{2}{l}{Binding} \\
    & $m_\uu$ & $m_\cc$ & $\epsilon_\uu^\text{M}$ & $\epsilon_\cc^\text{M}$ & $\alpha_\uu^\text{M}$ & $\alpha_\cc^\text{M}$ & $c^\text{M}$ & $s_\cc$ & $s_\uu$ & $\epsilon_\cc^\text{S}$ & $\epsilon_\uu^\text{S}$ & $\alpha_\cc^\text{S}$ & $\alpha_\uu^\text{S}$ & $c^\text{S}$ & $\epsilon_0$ & $\alpha$ \\ 
    \midrule
    $\mathcal{\theta}_2$ & $1.2$ & $2.71$ & $26.04$ & $20.72$ & $2.65$ & $2.63$ & $-0.10$ & $0.60$ & $3.0$ & $15.61$ & $57.08$ & $6.83$ & $1.08$ & $-0.06$ & $25$ & $5$ \\
    $\mathcal{\theta}_4$ & $1.11$ & $2.92$ & $26.05$ & $20.72$ & $2.69$ & $2.9$ & $-0.08$ & $0.5$ & $3.71$ & $15.69$ & $57.04$ & $6.84$ & $1.0$ & $2.13$ & $24.99$ & $4.98$ \\
    $\mathcal{\theta}_5$ & $1.24$ & $3.03$ & $25.63$ & $21.10$ & $2.99$ & $2.35$ & $0.66$ & $0.49$ & $3.81$ & $15.48$ & $46.0$ & $5.69$ & $1.13$ & $-0.69$ & $24.99$ & $4.99$ \\
    $\mathcal{\theta}_6$ & $1.19$ & $2.71$ & $26.04$ & $20.72$ & $2.65$ & $2.63$ & $-0.10$ & $0.6$ & $3.75$ & $15.61$ & $57.08$ & $6.83$ & $1.08$ & $-0.06$ & $25$ & $5$ \\
   \bottomrule
  \end{tabular}
  \caption{List of all model parameters of the different designs analyzed in Fig.~\ref{fig:si_measures}. The parameters describe the energy profiles of Machine and Source as well as the binding between them.}
  \label{tab:params_si}
\end{table*}

\subsection{Measures of merit for additional designs}

The measures of merit for different designs in addition to those analyzed in the main text are presented in Fig.~\ref{fig:si_measures}. The physical model parameters $\theta$ are listed in Table~\ref{tab:params_si}. Notice that the green designs here correspond to the orange design presented in Fig.~6 of the main text. Figure~\ref{fig:si_measures} demonstrates the robustness of the energy-delivery mechanism over a broad range of dissociation timescales and different designs as well as nontrivial tradeoffs between these measures when changing the energy profiles or $\tau_\dd$. For example, at $\tau_\mathrm{d}=100$, the blue design has the best performance, but the orange design shows significantly better power and efficiency. The relative importance of these quantities depends on the application, and in principle one can find energy profiles that maximize certain quantities over others.

As expected, the performance of the dark green design vanishes for large $\tau_{d}$ since the dissociation energy barrier of $25\,\kT$ is much larger than the Machine's internal barrier. However, when $\epsilon_0$ is lowered to $15\,\kT$, the performance plateaus around $p \approx 3$. We compare these designs to the previous ones without any time dependence in the binding energy (represented by $\tau_\dd = \infty$ in Fig.~\ref{fig:si_measures}). Crucially, the performance of the green design remains above 1, meaning that the energy-delivery mechanism works for intermediate $\epsilon$ without any forced dissociation. The number of Source recharging events plateaus for the light green designs when the binding energy effectively sets the dissociation timescale. The four designs with $\epsilon_0\approx25$ show a power-law decay for high $\tau_\dd$.

\section{Interpretation of performance}

\begin{figure*}[htb]
    \includegraphics[width=\textwidth]{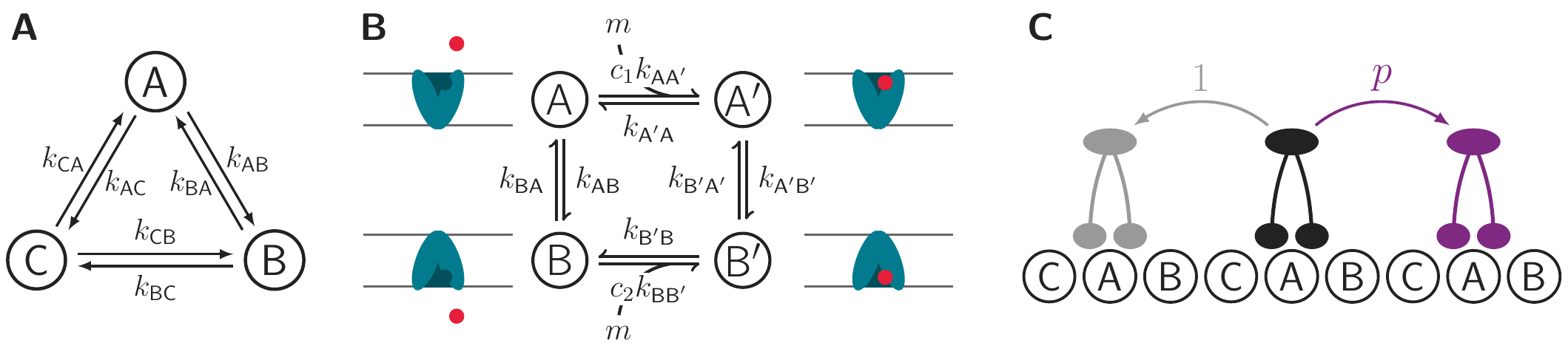}\centering
    \caption{Interpretation of performance. \textsf{(A)} A 3-state system cannot exhibit a net flux in equilibrium, but if one reaction is driven by our mechanism with a performance $p$, the system can experience roughly $p$ cycles in one direction for every 1 cycle in the opposite direction, enabling a net flux in steady state. \textsf{(B)} Synthetic pump. Without driving the pump, the concentrations of a molecule $m$ (red circle) are equal on both sides of the barrier, $c_2=c_1$. When using our energy-delivery mechanism, we can drive \eg the $A'\to B'$ reaction with a performance $p$ resulting in steady-state concentrations $c_2 = p\,c_1$, meaning we can increase the concentration of molecule $m$ on one side by a factor of $p$. \textsf{(C)} Synthetic walker. Like a Kinesin motor protein, the walker binds and moves along a filament, which is composed of three periodically repeating monomers. The walker binds to a monomer and can ``step'' either to the left or to the right. For every complete step the walker takes to the left (\ie~$A\rightarrow C\rightarrow B\rightarrow A$), it takes $p$ complete steps to the right (\ie~$A\rightarrow B\rightarrow C\rightarrow A$) when using the energy-delivery mechanism.}
    \label{fig:si_interpretation}
\end{figure*}

Figure~7 in the main text shows that the performance is slightly larger than, but close to, the increase in the number of actions of the Machine per unit time. For convenience, here we will assume that this relation is exact, though this means our conclusions may be a slight overestimation.

\subsection{Cyclic flux in a 3-state system}
Figure~\ref{fig:si_interpretation}\textsf{A} shows a 3-state system with all-to-all transitions. In equilibrium, detailed balance implies that $k_{AB}k_{BC}k_{CA} = k_{AC}k_{CB}k_{BA}$, so that the clockwise flux is equal to the counterclockwise flux:
\begin{equation}
    \frac{f_\mathrm{cw}}{f_\mathrm{ccw}} = \frac{k_{AB}k_{BC}k_{CA}}{k_{AC}k_{CB}k_{BA}} = 1.
\end{equation}
Now, consider driving the $A\rightarrow B$ reaction with our mechanism with a performance $p$. The $A\rightarrow B$ transition in the powered system will experience an effective rate of $k_{AB}^p = p\, k_{AB}$. Therefore, the net number of cycles will no longer be zero and 
\begin{equation}
    \frac{f^p_\mathrm{cw}}{f_\mathrm{ccw}} = \frac{k_{AB}^pk_{BC}k_{CA}}{k_{AC}k_{CB}k_{BA}} = p.
\end{equation}
Therefore, we see that a performance of $p$ means that a cyclic system can experience, \emph{e.g.}, up to $p$-times more clockwise cycles than counterclockwise cycles in steady state. We now explore what this implies in two practical examples.

\subsection{Example 1: A synthetic pump}
Consider a nanomachine that, like a protein pump, spans a barrier with the potential to transport a small molecule, $m$, across the barrier. As shown in Fig.~\ref{fig:si_interpretation}\textsf{B}, the nanomachine can transition between being open to one side of the barrier and the other. A molecule of $m$ can associate or dissociate only on the side to which the nanomachine is open. Figure~\ref{fig:si_interpretation}\textsf{B} shows the 4 relevant states and the 8 relevant transitions. If the transitions between $A\xrightleftharpoons[]{} B$ are unaffected by the presence of $m$ and dissociation is the same on both sides, then we have $k_{AB}=k_{A^\prime B^\prime}$, $k_{BA}=k_{B^\prime A^\prime}$, $k_{AA^\prime}=k_{BB^\prime}$, and $k_{A^\prime A}=k_{B^\prime B}$. In steady state, the clockwise flux must equal the counterclockwise flux, implying that $c_2 = c_1$: the concentration of $m$ is equal on both sides.

However, we can use our energy delivery mechanism to change this. For example, if we drive the $A^\prime \rightarrow B^\prime$ reaction with a performance $p$, then we have $k_{A^\prime B^\prime}=p\,k_{AB}$. Alternatively, if we drive the $B^\prime \rightarrow B$ dissociation with a performance $p$, then we have $k_{B^\prime B} = p\,k_{A^\prime A}$. In both cases, the steady-state concentrations are $c_2 = p\,c_1$, meaning we can increase the concentration of $m$ on one side by a factor of $p$.

\subsection{Example 2: A synthetic walker}
Consider a nanomachine that, like a Kinesin motor protein, binds and moves along a filament. 
As shown in Fig.~\ref{fig:si_interpretation}\textsf{C}, the filament is composed of three monomers that periodically repeat: $A-B-C-A$, etc. The walker binds to a monomer and can ``step'' either to the left or the right. Thus, the reaction network is exactly that shown in Fig.~\ref{fig:si_interpretation}\textsf{A}, and the results are the same: for every complete step the walker takes to the left (\ie~$A\rightarrow C\rightarrow B\rightarrow A$), it takes $p$ complete steps to the right (\ie~$A\rightarrow B\rightarrow C\rightarrow A$) if one of the reactions is powered by our mechanism.

Importantly, in neither of these examples have we explained \emph{how} our mechanism would be incorporated into the geometry of a nanomachine to drive the desired reactions. Understanding this requires a microscopic design and model for these nanomachines that goes well beyond the scope of the present work. Rather, we have shown what can be achieved in principle with an energy-delivery mechanism that operates with a performance $p$.

\end{document}